\documentclass{aa}  
\usepackage{hyperref}
\usepackage{placeins}
\usepackage{orcidlink}

\usepackage{xcolor}
\hypersetup{    
    colorlinks=true,
    citecolor = blue,
    linkcolor=blue,
    filecolor=magenta,      
    urlcolor=cyan,
}
\usepackage{multirow}
\usepackage{tikz} 

\usepackage{graphicx}
\usepackage{txfonts}

 \def\kms{\, \rm km\,s^{-1}}
 
 \def\Msun{\,\mathrm{M}_\odot}
 \def\Rsun{\,{\rm R}_\odot}
 \def\Lsun{\,{\rm L}_\odot}

\newcommand{\Case}[1]{Case\,#1}

 \def\qi{q_\mathrm i}
 
 \def\days{\,\text{d}}

 \def\erg{\,\text{erg}}
 \def\ergs{\,\text{erg}\,\mathrm{s}^{-1}}
 \def\tauinfl{\tau_\mathrm{infl}}

 \def\Mej{M_\mathrm{ej}}
 
 \newcommand{\Type}[1]{\text{Type}\,\text{#1}}

 \newcommand{\multilinecomment}[1]{}
\newcommand{\yr}{\,\mathrm{yr}}

\def\simle{\mathrel{\hbox{\rlap{\hbox{\lower4pt\hbox{$\sim$}}}\hbox{$<$}}}}
\def\simgr{\mathrel{\hbox{\rlap{\hbox{\lower4pt\hbox{$\sim$}}}\hbox{$>$}}}}

\newcommand{\EDIT}[1]{{#1}}
\newcommand{\EDITTWO}[1]{{#1}}

\defcitealias{Ercolino25_popsynth_CCSNE_I}{E26}
\def\PI{\citetalias{Ercolino25_popsynth_CCSNE_I}}
\defcitealias{Valli25_BeXrayBinaries_kicks}{V25}
\def\Valli{\citetalias{Valli25_BeXrayBinaries_kicks}}
\defcitealias{MHLC16}{M16}

\defcitealias{DM25_kicks_isolated_ns}{DM25}
\def\DM{\citetalias{DM25_kicks_isolated_ns}}
\defcitealias{Hobbs2005_kicks_isolated_pulsars}{H05}

\defcitealias{Kruckow18_COMBINE}{K18}
\def\COMBINE{\citetalias{Kruckow18_COMBINE}}
\def\TZO{T$\dot{\mathrm{Z}}$O}
\newcommand{\SN}[1]{\object{SN{#1}}}

\begin{document}

   \title{Neutron star-companion interaction in core collapse supernovae}
   \subtitle{Population synthesis based on detailed binary evolution models.}

   \author{Andrea Ercolino
          \inst{1}\thanks{Corresponding author. \email{aercolino@astro.uni-bonn.de}}\orcidlink{0000-0002-2807-5253}
          \and Norbert Langer\inst{1,2}\orcidlink{0000-0003-3026-0367}
          \and Avishay Gal-Yam \inst{3}\orcidlink{0000-0002-3653-5598}
          \and Abel Schootemeijer \inst{1}\orcidlink{0000-0002-2715-7484}
          \and Caroline Mannes\inst{1,2}\orcidlink{http://orcid.org/0009-0002-4207-6348} 
          \and \\
          Harim Jin \inst{1,4}\orcidlink{0000-0003-2946-9390}
          \and Ruggero Valli\inst{4}\orcidlink{0000-0003-3456-3349}
          \and Selma de Mink\inst{4}\orcidlink{0000-0001-9336-2825}
          \and Luc Dessart\inst{5,6,7}\orcidlink{0000-0003-0599-8407}
}

   \institute{Argelander Institut für Astronomie,
              Auf dem Hügel 71, DE-53121 Bonn, Germany
         \and
          Max-Planck-Institut für Radioastronomie, Auf dem Hügel 69, DE-53121 Bonn, Germany
          \and
            Department of Particle Physics and Astrophysics, Weizmann
Institute of Science, 234 Herzl Street, IL-7610001 Rehovot, Israel          \and
            Max-Planck-Institut für Astrophysik, Karl-Schwarzschild-Straße 1, DE-85748 Garching bei München, Germany
    \and 
     Institut d'Astrophysique de Paris, CNRS-Sorbonne Universit\'e, 98 bis boulevard Arago, F-75014 Paris, France 
     \and 
     French-Chilean Laboratory for Astronomy, IRL 3386, CNRS
     \and 
     Pontificia Universidad Católica de Chile, Casilla 306, Santiago, Chile
     }
   \date{Received Month Day, 2026; accepted Month Day, 2026}

 
  \abstract
{Most massive stars live in binary systems. When the first supernova in a binary occurs, the ejecta hit the companion, which may inflate as a consequence, and then interact with the newly formed compact object. 
The recent Type Ic supernova \SN{2022jli} shows a periodic modulation in its emission, which is interpreted as evidence for such interaction.}
{We derive predictions for the occurrence rate and observables of supernovae exhibiting these companion - compact-object interactions (CCIs).}
{We analyze a comprehensive, state-of-the-art grid of detailed binary stellar evolution models, and implement analytic prescriptions for the expansion of the companion star following its interaction with the supernova ejecta. We then employ the newly developed population synthesis code \texttt{SN-ORACLE} to derive the distribution functions of the properties of the supernovae affected by CCI and their companions, where we use different explodability and neutron star birth kick distributions.}
{We find that periodic CCI is expected to occur in more than half of the binary systems that produce a hydrogen-poor core collapse supernova and are not disrupted, while the occurrence rate in systems producing hydrogen-rich supernovae is small. 
We find broad period ranges, peaking around $20-50\days$, with the interaction lasting for $0.5-10\yr$.
We identify specific binary evolution models that reproduce the observed period of the light curve undulations of \SN{2022jli}, \SN{2015ap}, and \SN{2022esa}. 
The inflation of the companion also increases their luminosity and brightness, increasing their detectability with current instruments. For \SN{2022jli}, our best fitting models predict a $J$-band magnitude of $21-23$ for up to $\sim10\yr$.}
{We find that up to $\sim27\%$ of H-poor supernovae could show periodicity in their light curves, while only a few such events have been identified so far. 
Our results may help find periodic CCI features in future and archival supernova observations. } 

   \maketitle
%

\section{Introduction}

Supernova light curves and spectra constrain the properties of their massive star progenitors \citep{Blinnikov1993_STELLA,Smith2011_obsSNfractions, Dessart2016_IIb_Ibc_RadTransf2, Morozova15_LCofCCSNe_SNEC, GalYam17_SN_Book,  Williamson21_1994I_Ic_with_He}. 
The presence of a binary companion is expected to affect the evolution of most massive stars \citep{Sana_massive_stars_binaries,Sana14_Ostar_binaries,Almeida17_FLAMES_OB_binaries,MoeDiStefano2017,Sana2025_BLOEM_binaryfraction_Z}, and the properties of the resulting supernovae \citep{Smith2011_obsSNfractions, Langer_review_2012, Eldridge13_deathmassivestarII_Ibc, Zapartas19_howmany_Hrich_Sne_from_binaries, Ercolino25_popsynth_CCSNE_I}. However, direct detections of surviving companion stars to the progenitors of observed supernovae are rare, with only a few identified through long-term post-explosion photometric monitoring that revealed a remaining luminous star at the explosion site (e.g., \object{SN1993J}, \citealt{Maund2004_1993J, Fox14_companion1993J}; \object{SN2001ig}, \citealt{Ryder2018_companion2001ig}; \object{SN2006jc}, \citealt{Sun2020_companions2006jc_2015G}; \object{SN2011dh}, \citealt{Maund2019_companion2011dh}; \object{SN2013ge}, \citealt{Fox2022_companion_2013ge}) as well as pre-explosion photometry (e.g., \object{SN2019yvr}, \citealt{Sun2022_2019yvr_companion}).

In recent years, a growing number of supernovae have been observed with undulations or bumps in their light curves, especially in some H-poor superluminous supernovae (SLSNe-I, \citealt{Nicholl2016_2015bn_SLSN_undulation, West_2020qlb_SLSNI_oscillations, Hosseinzadeh22_SLSNe_bumps_and_oscillations, Chen2023_SLSNEI_CSM_Magnetar_undulations,  Kumar25_2024afav_SLSNI_with_undulations, Farah25_SLSNI_undulations_magnetar}), H-rich SLSNe-II (e.g., \object{iPTF14hls}, \citealt{Arcavi17_iPTF14hls}), \Type IIn supernovae (e.g., \object{iPTF13z}, \citealt{Nyholm17_iPTF13z_IIn_with_undulations_bump}), and the so-called ``bactrian'' supernovae, \citep{Kuncarayakti2023_SN2022xxf_IcBL_multipeak}. These features may arise from additional powering mechanisms, including interaction with circumstellar material \citep{Moriya11_CSM_SNe_RSG,Chatzopoulos12_CSM_SLSN} or a central engine (e.g. magnetar spin-down, \citealt{Maeda2007_SN2005bf_Ib_magnetarspindown, KasenBildstein2010_MagnetarSpindown_poweringSNe,Dessart2018_iPTF14hls,Chen2023_SLSNEI_CSM_Magnetar_undulations, Farah25_SLSNI_undulations_magnetar}, or fall-back accretion, \citealt{DexterKasen13_FallbackAccretion_SNe}). 

In the remarkable \Type Ic supernova \object{SN2022jli} \citep{Moore2023_SN2022jli_Ic_Binary, Chen2024_SN2022jli_binary}, a periodic signal is instead attributed to the presence of a non-degenerate stellar companion \citep{Moore2023_SN2022jli_Ic_Binary, Chen2024_SN2022jli_binary, Zhang25_2022jli_gammarays}. This comes from the concurrence of three features from this supernova that display modulation with the same period of $12.4\days$: (i) the optical flux \citep{Moore2023_SN2022jli_Ic_Binary}, (ii) the velocity and intensity of the emission-line H$\alpha$ \citep{Chen2024_SN2022jli_binary}, and (iii) the $\gamma$-ray flux \citep{Zhang25_2022jli_gammarays}. Additionally, this supernova has a second peak in the light curve, which preceded the onset of the undulations, and is thought to be powered by magnetar spin-down \citep{Orellana25_2022jli_magnetar, Cartier26_2022jli_magnetar}. Consequently, \SN{2022jli} provides a unique laboratory for probing and constraining theoretical models of binary evolution and core-collapse supernova explosions. 

\begin{figure}
    \centering
    \includegraphics[width=0.95\linewidth]{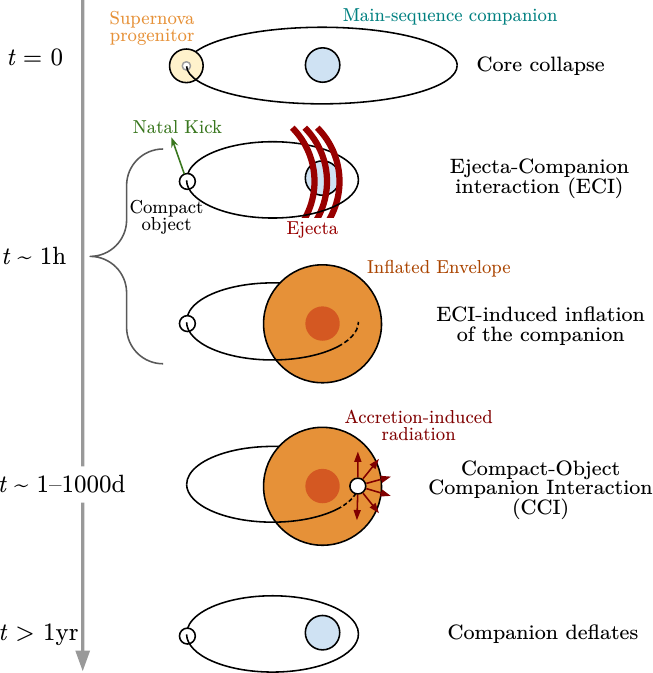}
    \caption{Schematic evolution of SN2022jli-like transients in binaries, from the explosion (top), through ejecta-companion interaction (ECI) and compact-object-companion interaction (CCI), to the deflation of the companion (bottom). Characteristic timescales are shown on the left, with $t=0$ marking core collapse.}
    \label{fig:cartoon}
\end{figure}

\citet{Hirai25_2022jli_ECI} showed that periodic interactions between the compact object and the shock-heated companion's envelope (which we refer to as compact-object companion interaction; CCI) plausibly explain the observed properties of \object{SN2022jli} (see also \citealt{Cary26_CCI_spindown_NS} and \citealt{Lu25_CCI}), and we illustrate this in Fig.\,\ref{fig:cartoon}. After the supernova explosion, part of the ejecta collides with the companion star, the so-called ejecta-companion interaction (ECI), which can include observational features on both the light curve \citep{Kasen10_ejecta_ccsn_impact_sec} and spectra \citep{Dessart2020_spectraIa} of the supernova. This interaction has been widely studied for both \Type Ia supernovae \citep{Pakmor08_Ia_ejecta_vs_ms, Marietta2000_Ia_comapnion_ms_2DHD, Liu15_UV_Ia_signatures_companion, Dessart2020_spectraIa} as well as core-collapse supernovae \citep{Kasen10_ejecta_ccsn_impact_sec, Hirai14_CCSNe_ejecta_impact_companion, Liu15_ccsne_ejecta_interaction_with_ms, HiraiYamada15_ECI}. 
ECI can also ablate and strip material from the companion \citep{Wheeler75_SNe_companion_ablation, Hirai14_CCSNe_ejecta_impact_companion, TaamFryxell84_sne_ejecta_companion, Chen23_ECI_models} and deposit enough energy to produce an extended low-density envelope. Depending on the intercepted energy, this inflated phase can last for years after the explosion \citep{Hirai23_ECI, Chen23_ECI_models, Lu25_CCI}.

When the companion star is strongly inflated, the newly formed compact object, either a neutron star (NS) or a black hole (BH), may interact with it, emitting radiation periodically. Whether and how this interaction occurs is mainly determined by the natal kick of the compact object, whose origin and distribution remain active research topics \citep[for a recent review, see][]{Popov25_natalkicks_review}.

In this study, our objective is to produce theoretical predictions for the number and features of the core-collapse supernovae that undergo the interaction process proposed by \cite{Hirai25_2022jli_ECI} for \SN{2022jli}, where the newly formed compact object interacts with its perturbed main-sequence companion right after the supernova explosion. In Sect.\,\ref{sec:methods}, we describe the   population-synthesis code adopted, originally developed in \cite{Ercolino25_popsynth_CCSNE_I}, hereafter \PI. In Sect.\,\ref{sec:results}, we present our main results, which we compare to the observed sample of supernovae with light curve undulations in Sect.\,\ref{sec:observations}. Theoretical uncertainties of our model are discussed in Sect.\,\ref{sec:discussion}, and we summarize our findings in Sect.\,\ref{sec:conclusions}.

\FloatBarrier

\section{Method} \label{sec:methods}
We used the population-synthesis code \texttt{SN-ORACLE}\footnote{v0.2, \url{https://github.com/andreaerk/SN-ORACLE}, \url{https://doi.org/10.5281/zenodo.23105688}
} developed in \PI
, to predict the relative fractions and properties of different supernova types. The code uses a large comprehensive grid of  \texttt{MESA} \citep{MESA_I, MESA_II, MESA_III, MESA_IV} single and binary stellar evolutionary models at galactic metallicity (\citealt{Jin2024_boron, Jin25_BonnGal}, \PI). Below, we describe the additional physical prescriptions implemented here: (i) natal kicks, (ii) the companion's response to ECI, and (iii)  interaction between the newborn NS and the ECI-inflated companion immediately after explosion.

We adopted the same assumptions as in \PI\ and summarize only those most relevant for this study. We assumed that successful explosions produced NSs, whereas failed explosions do not eject significant material to impact the companion and directly collapse to form BHs that do not receive natal kicks (see Sect.\,2.2.3 in \PI). Consequently, systems that form BHs cannot produce SN\,2022jli-like events and are excluded from this analysis (but see Sect.\,\ref{sec:discussion:additionalpathways}).

\EDIT{For each binary evolution model in the grid, we assigned explosion parameters (ejecta mass, remnant mass, and explosion energy) from core-collapse single-star or single He-star models (depending on the mass-transfer history; \citealt{Schneider_preSNevo_stripped_2021}) with the closest CO-core mass. These parameters were derived using the semi-analytical explosion model of \cite{MHLC16} (see \PI\ for details).}

\subsection{Neutron-star kicks}\label{sec:method:kicks}
Following the explosion of the primary star in a binary, we assigned a natal kick to the newly born NS by drawing its magnitude and direction randomly from prescribed probability distributions. This natal kick is added to the \cite{BlaauwKick_1961} kick from the instantaneous mass loss of the ejecta. We adopted the following kick distributions:
\begin{itemize}
    \item \textbf{\DM} \citep{DM25_kicks_isolated_ns}: the kick distribution is calibrated on the observed velocities of isolated pulsars and constitutes a revised version of that by \cite{Hobbs2005_kicks_isolated_pulsars}.
\item \textbf{\COMBINE} \citep{Kruckow18_COMBINE}: the kick prescription distinguishes three sub-populations: (i) a low-kick component from doubly-stripped progenitors; (ii) a medium-kick component from progenitors that underwent a single stripping episode in binaries \citep{TaurisBailes96_ms_pulsars_sesne_lowkicks,Tauris17_DNS_lowkicks}; and (iii) a high-kick component from non-interacting (effectively single) progenitors \citep{Hobbs2005_kicks_isolated_pulsars}.   
\item \textbf{\Valli} \citep{Valli25_BeXrayBinaries_kicks}: the kick distribution is proposed to explain observed Be X-ray binaries, and it comprises three sub-populations, which we categorize analogously to \COMBINE, but with distinct distributions.
\end{itemize}
Table\,\ref{tab:kicks} details the direction and magnitude distributions of the different kick prescriptions.
\begin{table*}[]
\centering
\caption{Details on the different kick distributions adopted in our population synthesis models.}
\begin{tabular}{rllll}
Prescription & Magnitude  & Angle   & Affected    \\
  & distribution(s) & distribution(s) & models &   \\\hline\hline
\DM & $\mathrm{Lognormal}(5.6,0.58) \ [\mathrm{km}\,\mathrm{s}^{-1}]$  & Isotropic & All\\[0.3em]
\COMBINE $\left\{ \begin{tabular}[c]{@{}l@{}}\\ \\ \\ \end{tabular} \right.$ & \begin{tabular}[c]{@{}l@{}}$\mathcal{MB}(265\kms)$\\ $0.8\cdot\mathcal{MB}(120\kms)+0.2\cdot\mathcal{MB}(200\kms)$\\ $0.8\cdot\mathcal{MB}(60\kms)+0.2\cdot\mathcal{MB}(200\kms)$\end{tabular} & \begin{tabular}[c]{@{}l@{}}Isotropic\\ Isotropic\\ Isotropic\end{tabular}       & \begin{tabular}[c]{@{}l@{}}Effectively Single\\ Underwent \Case A/B/C RLOF\\ Underwent Case BB RLOF\end{tabular}  \\[1.6em]
\Valli $\left\{ \begin{tabular}[c]{@{}l@{}}\\ \\ \\ \end{tabular} \right.$ & \begin{tabular}[c]{@{}l@{}}$\mathcal{MB}(300\kms)$\\ $\mathcal{N}(100\kms,11\kms)$\\ $\mathcal{MB}(5\kms)$\end{tabular} & \begin{tabular}[c]{@{}l@{}}Isotropic\\ $5^\circ$ from polar axis\\ Isotropic\end{tabular} & \begin{tabular}[c]{@{}l@{}}Effectively Single\\ Underwent \Case A/B/C RLOF \\ Underwent Case BB RLOF\end{tabular} \\[1.6em]
\end{tabular}
\tablefoot{The probability distribution notation is as follows: $\mathcal{MB}(\sigma)$ denotes a Maxwell-Boltzmann distribution with scale parameter $\sigma$; $\mathrm{Lognormal}(\mu,\sigma)$ denotes a log-normal distribution with logarithmic mean $\mu$ and logarithmic standard deviation $\sigma$; and $\mathcal{N}(\mu,\sigma)$ denotes a normal distribution with mean $\mu$ and standard deviation $\sigma$.
}
\label{tab:kicks}
\end{table*}

For each evolutionary model in the binary grid, we sampled the kick magnitude and direction $10^3$ times using the chosen prescription. We assumed that the \EDIT{long-term} post-supernova binary evolution depends only on the radius evolution of the non-degenerate companion and the post-kick periastron distance of the newly formed compact object, and that the orbit circularizes with a semi-major axis equal to this periastron distance. We assumed that any future episodes of mass-transfer via Roche-lobe overflow (RLOF) \EDIT{while the donor star is in thermal equilibrium} are dynamically unstable: if the donor is on the main sequence, it is disrupted after briefly becoming a \cite{ThorneZytkow_77} object (\TZO); if the donor is more evolved, the system enters a common-envelope phase that we assume ejects the H-rich envelope, leaving a He-star + NS binary. We did not model beyond the second supernova.

\subsection{ECI: Ejecta-companion interaction}\label{sec:method:ECI}
Following the results from \cite{Hirai18_ECI}, \cite{Ogata21_ECI}, and \cite{Hirai23_ECI}, ECI has been shown to induce a substantial radial expansion of the companion. Here, we adopted their prescriptions and implemented ECI with analytic formulae, which we summarize below.

The key parameter governing ECI is the fraction of the kinetic energy of the supernova ejecta $E_\mathrm{kin,ej}$ intercepted by the companion star, $E_\mathrm{heat}$, which is given by
\begin{equation}\label{eq:Eheat}
E_\mathrm{heat} = p\cdot E_\mathrm{kin,ej} \cdot \tilde\Omega_\mathrm{eff}    ,
\end{equation}
with $p=0.08$ an efficiency factor \citep{Hirai18_ECI}. Here,  $\tilde\Omega_\mathrm{eff}$ is the fractional solid angle subtended by the companion as seen from the exploding progenitor, which is a function of the companion's radius $R_2$ and its separation $a$: 
\begin{equation}\label{eq:OmegaEff}
    \tilde\Omega_\mathrm{eff}=\frac{\Omega(R,a)}{4\pi}=\frac 1 2 \left(1-\sqrt{1-\left(\frac{R_2}{a}\right)^2}\right).
\end{equation}

The intersected energy $E_\mathrm{heat}$ drives the stellar expansion until the luminosity matches the local Eddington luminosity in the layers below the surface convective region \citep{Ogata21_ECI}, given by
\begin{equation}\label{eq:Lmax}
L_\mathrm{max} = \frac{4\pi Gc M_2}{\kappa_\mathrm{fit}(M_2)}    ,
\end{equation}
where $\kappa_\mathrm{fit}(M_2)=1.24(1-0.02 \,M_2/\Msun)\mathrm{cm}^2\,\mathrm{g}^{-1}$ represents the empirically fitted local opacity in this layer, calibrated for stars of different companion masses $M_2$ in the range $3-20\,\Msun$ \citep{Ogata21_ECI}.
The radial expansion of the star is described by the empirical relation 
\begin{equation}\label{eq:Rmax}
    \log_{10} \frac{R_\mathrm{max}}{\Rsun} = -\frac{1}{14}\left(\log_{10}\frac{E_\mathrm{heat}}{8\times 10^{49}\erg}\right)^2+3.1,
\end{equation}
which should be taken as an upper limit \citep{Ogata21_ECI}. The duration of this inflated phase $\tauinfl$ can be estimated given the total injected energy, $E_\mathrm{heat}$, and the rate at which the star radiates this energy, $L_\mathrm{max}$, leading to the relation
\begin{equation}\label{eq:tau_ECI}
\tauinfl=\alpha \frac{E_\mathrm{heat}}{L_\mathrm{max}},
\end{equation}
with $\alpha\sim0.18$ an efficiency factor \citep{Ogata21_ECI}. The luminosity $L_\mathrm{max}$ and radius $R_\mathrm{max}$ of the inflated companion are found to remain roughly constant for a time $\tauinfl$ after the onset of ECI (see Fig.3 in \citealt{Ogata21_ECI}). The ECI-driven expansion occurs on a dynamical timescale (\citealt{Hirai25_2022jli_ECI}, although it may be longer, see \citealt{Chen23_ECI_models, Lu25_CCI}) typically hours for main-sequence stars. Because this is much shorter than typical post-explosion orbital periods, we treat the expansion as instantaneous. We also assume ECI does not significantly ablate mass from the companion \citep{Hirai14_CCSNe_ejecta_impact_companion, Hirai18_ECI}, so its post-interaction mass remains roughly unchanged.

\subsection{CCI: Companion + Compact-object Interaction}\label{sec:method:CCI}
We interpret the periodic modulation observed in the light curve,  H$\alpha$ line and $\gamma$-ray flux of \SN{2022jli} as resulting from interaction between the newly born NS and the shock-inflated envelope of the (H-rich) companion star after ECI (see also \citealt{HiraiPosdi22_ECI_NS,Hirai25_2022jli_ECI, Cary26_CCI_spindown_NS, Lu25_CCI}). 

For CCI to occur, we require that the post-explosion periastron distance $r_\mathrm{peri}$ satisfies
\begin{equation}\label{eq:r_CCI}
    R_2 < r_\mathrm{peri} \leq R_{\mathrm{max}},
\end{equation}
where $R_2$ is the companion’s radius before ECI, and $R_\mathrm{max}$ its maximum radius following ECI. If Eq.\,\ref{eq:r_CCI} is satisfied, the NS only penetrates the companion’s low-density, shock-extended envelope, where we assume the resulting accretion-powered luminosity becomes observable as light curve modulations (see Sect.\,\ref{sec:disc:ECI_CCI:CCI}). Because the accreted material is relatively tenuous, the NS is not significantly decelerated, and we assume that the orbital parameters remain roughly unchanged after each periastron passage (\citealt{Hirai25_2022jli_ECI}). 

\EDIT{If $r_\mathrm{peri}>R_\mathrm{max}$, the NS completely avoids the companion and no CCI ensues. In the cases where $r_\mathrm{peri}\leq R_2$, the NS skims denser material, enhancing orbital energy dissipation. We assume this leads the NS and companion to form a \TZO\ or producing another terminal explosion, termed merger-burst \citep{Chevalier2012_mergerburst, Brennan2025_SN2022mop}. The lower limit on the allowed periastron distance is likely underestimated, since the star can fill its (eccentricity-corrected) Roche lobe when $R_2 \lesssim r_\mathrm{peri}/2$ and trigger mass transfer, which would likely be unstable. However, this phase could still be stable in some cases, and a merger may not always ensue, even if $r_\mathrm{peri}<R_2$ \citep[see the unbound penetration case in][]{HiraiPosdi22_ECI_NS}.} 

We also require that the time of the first periastron passage $t_\mathrm{1st-peri}$ occurs before the companion deflates: $t_\mathrm{1st-peri} \leq \tauinfl$, otherwise the NS would not be able to skim its shock-inflated envelope.
If the orbit remains bound and the \EDIT{post-explosion orbital period} $P_\mathrm{post-SN}$ is short enough, the NS can penetrate the companion's envelope multiple times. These episodes should produce periodic undulations in the supernova light curve with the same period as $P_\mathrm{post-SN}$, lasting for $\tauinfl$ (but see Sect.\,\ref{sec:disc:ECI_CCI}), and we can estimate how many of them occur as
\begin{equation}\label{eq:Ncci}
    N_\mathrm{inter} = 1+ \left\lfloor\frac{\tauinfl-t_\mathrm{1st-peri}}{P_\mathrm{post-SN}}\right\rfloor,
\end{equation}
where $\lfloor \cdot \rfloor$ is the floor function. A supernova is classified as affected by periodic CCI when $N_\mathrm{inter}\geq2$.

\section{Results}\label{sec:results}

Here, we present the number and properties of transients showing CCI in our population models. In. Sect.\,\ref{sec:results:example} we show results for the reference population model, and in Sect.\,\ref{sec:results:multiple} those from alternative models.

\subsection{Reference population model}\label{sec:results:example}
We use the fiducial population model from \PI, which assumes a 75\% birth-binary fraction, a \cite{Salpeter_IMF_55} initial mass function (IMF), and flat initial orbital-period $\log P_\mathrm{i}$ and mass-ratio $\qi=M_\mathrm{2,i}/M_\mathrm{1,i}$ distributions. Among all population models explored in \PI, the fiducial model produces the fewest binary mergers while maximizing the number of explosions (aside from the non-physical case with no BH formation), thus providing the largest parameter space for CCI. Following the first supernova, we apply the natal NS kick distribution from \Valli, which is the one that produces the closest number of NSs that remain bound in binaries after the first explosion compared to \citet[see Appendix\,\ref{sec:appendix:OB_cal}]{Schuermann25_SMC_pop_combine}. We refer to this as our reference model.

The reference population model shows no significant differences in the population-wide properties of core-collapse supernovae relative to the fiducial model in \PI, where the kicks were assumed to always disrupted binaries. The only notable change is a $4.0\%$ increase (decrease) in the fraction of \Type IIP/L (\Type Ibc) supernovae: as some binaries now remain bound after the first supernova, their companion can undergo RLOF with the NS and lose their H-rich envelopes (Sect.\,\ref{sec:method:kicks}), leading them to explode preferentially as \Type Ibc supernovae.

\subsubsection{Which supernovae types are affected by CCI?}\label{sec:example:CCI_numbers}

About $13\%$ of H-rich supernovae (\Type IIP/L, IIb, IIn) have a non-degenerate (typically main-sequence) companion at the time of explosion (cf. Fig.\,1 in \PI). This low fraction arises because these transients are \EDIT{mainly} produced by single stars, merger products, or secondary stars in binaries (i.e., the initially less-massive component). H-rich supernovae with a non-degenerate companion instead originate from the explosion of primary stars (i.e. the initially more-massive component) in wide binaries (Fig.\,\ref{fig:logPq}, Fig.\,D.1 in \PI), which are usually disrupted after the explosion. Consequently, CCI is extremely rare in H-rich supernovae, arising only $\sim0.01\%$ of them. Because of this, we exclude H-rich supernovae from further analysis.

\begin{figure}
    \centering
    \includegraphics[width=.71\linewidth]{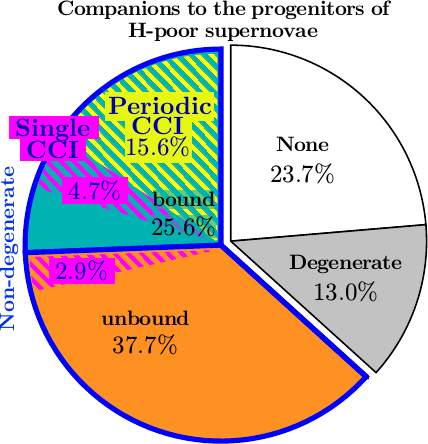}
\caption{Fraction of the companions of \Type Ibc and \Type Ibn supernova progenitors at the time of explosion in the reference population model (Sect.\,\ref{sec:results:example}). We distinguish those with no companion (single stars, merger products, or disrupted binaries; white), a degenerate companion (gray), or a non-degenerate companion (blue outline). Non-degenerate companions are further divided into those that remain bound (teal) and those that become unbound (orange) during the explosion. Models expected to undergo periodic CCI are highlighted with hatching, \EDIT{specifically if periodic (yellow) or just once (magenta).}}
     \label{fig:pies}
\end{figure}

For H-poor supernovae, about two-thirds of progenitors have a main-sequence companion at the time of explosion (of which fewer than half remain bound; Fig.\,\ref{fig:pies}) the rest have either a bound degenerate companion ($13\%$) or no companion ($\sim24\%$). They mainly originate in close binaries that have undergone mass transfer, which means they also have weaker kicks (Table\,\ref{tab:kicks}). Together, this allows a larger fraction of H-poor supernovae that undergo CCI (see Appendix\,\ref{app:parameter_space}). We find that $15.6\%$ of H-poor supernovae show periodic CCI, about half the fraction expected to retain a bound non-degenerate companion after explosion (Fig.\,\ref{fig:pies}). About \EDIT{$7.6\%$} show only a single CCI event, \EDIT{mostly from binaries that remain bound (4.7\%), with the rest from disrupted systems (2.9\%).}

A subset of progenitors of H-poor supernovae deserves further attention. These are the stripped stars undergoing \Case BB RLOF, which are identified in \PI\ as progenitors of \Type Ibn supernovae. They make up $\sim7\%$ of all H-poor supernovae and always have a non-degenerate companion at the time of explosion, since we neglect the post–common-envelope channel with a stripped star and a NS (see \PI). Because of the very small kicks featured in these supernovae (Table \ref{tab:kicks}), the newly born NS remains bound with the non-degenerate companion after the explosion. As a result, about one third of the supernovae produced from these progenitors undergo periodic CCI (which make up about a sixth of all H-poor supernovae with periodic CCI).

\subsubsection{Properties of the undulations produced by CCI}\label{sec:example:CCI_properties}

\begin{figure}
\centering
    \includegraphics[width=0.91\linewidth]{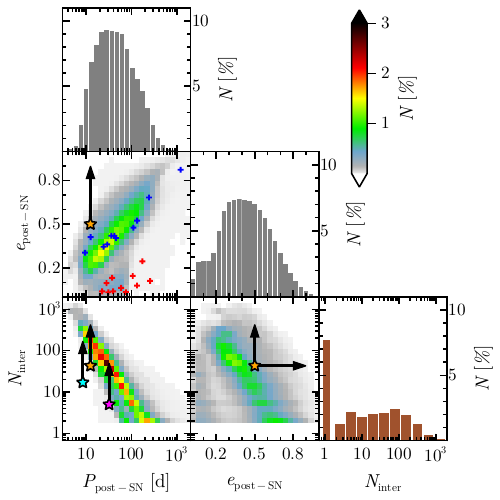}
\caption{Population-wide distributions of post-explosion orbital period $P_\mathrm{post-SN}$, eccentricity $e_\mathrm{post-SN}$, and the number of times the NS penetrates the companion's envelope $N_\mathrm{inter}$, for H-poor supernovae with periodic CCI in the reference model. All histograms and 2D plots are normalized to the total number of supernovae with periodic CCI, except the $N_\mathrm{inter}$ histogram (brown), which includes cases with $N_\mathrm{inter}=1$ and is normalized to the total number of H-poor supernovae. Star symbols show inferred values for SN\,2022jli (orange), SN\,2015ap (cyan), and SN\,2022esa (magenta; see Table\,\ref{tab:inferred_data_2022jli}). The $P_\mathrm{post-SN}$–$e_\mathrm{post-SN}$ panel includes Be/X-ray binaries compiled by \Valli, split into a low-$e_\mathrm{post-SN}$ (red) and high-$e_\mathrm{post-SN}$ (blue) group.}
\label{fig:cornerplot_PeN}
\end{figure}

Figure\,\ref{fig:cornerplot_PeN} reports the distribution of $P_\mathrm{post-SN}$, $e_\mathrm{post-SN}$ and $N_\mathrm{inter}$ in the binaries that produce supernovae exhibiting periodic CCI. These quantities respectively translate to the undulation period in the light curve (as well as of the other observable like the H$\alpha$ line's shift and $\gamma$-ray flux), the skewness of the undulation, and the number of undulations expected.  

The supernovae with periodic CCI have post-supernova orbital periods\footnote{All population-wide ranges correspond to the 5th-95th percentiles of the underlying distributions.} $P_\mathrm{post-SN}=9.8-226\days$, peaking at $20-70\days$. The orbital eccentricities peak around $e_\mathrm{post-SN}\sim0.4$. We identify two groups: one where the eccentricity increases with orbital period and another with low eccentricities $\simle0.20$. This dichotomy is expected when adopting the kicks of \Valli, with the low-eccentricity group arising from binary evolution models that undergo \Case BB RLOF before core-collapse.

In about one-third of supernovae with periodic CCI, the NS interacts with the companion's inflated envelope fewer than ten times ($N_\mathrm{inter}\leq 10$), while in another third $>50$. Larger $N_\mathrm{inter}$ occurs in models with \EDIT{shorter} post-explosion orbital periods (Eq.\,\ref{eq:Ncci}, Fig.\,\ref{fig:cornerplot_PeN}). Although supernovae with $N_\mathrm{inter}=1$ lack periodic modulations, they may still display notable features (Sect.\,\ref{sec:observations:other_transients}).

\subsubsection{Companion properties}\label{sec:res:example:companion}

\begin{figure}
    \centering
    \includegraphics[width=.88\linewidth]{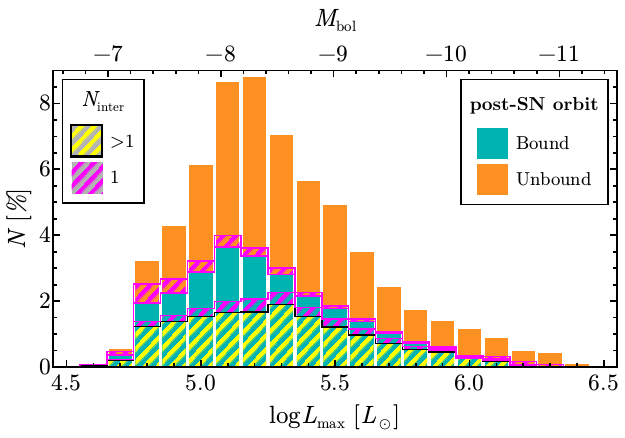}
    \includegraphics[width=.88\linewidth]{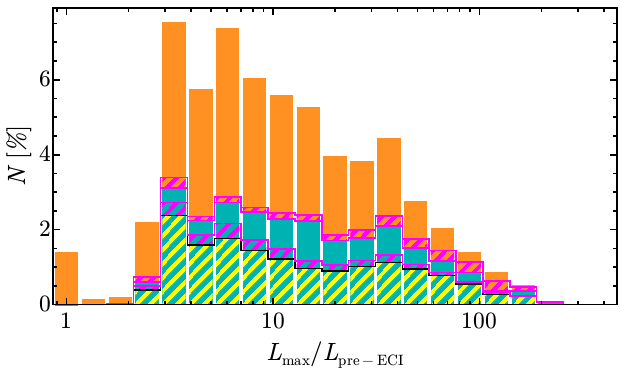}
\caption{Distribution of the maximum luminosity $L_\mathrm{max}$ (top) of ECI-inflated non-degenerate companions to all H-poor supernovae in the reference population model. The bottom panel shows the companion’s maximum luminosity relative to the its pre-ECI luminosity. The histograms are normalized to the total number of H-poor supernovae. Orange and teal filling show systems with unbound and bound companions after explosion, respectively. Hatching marks those exhibiting periodic ($N_\mathrm{inter}>1$, yellow), and single CCI ($N_\mathrm{inter}=1$, magenta) respectively.}
     \label{fig:Lmax}
\end{figure}

In supernovae exhibiting CCI, companion stars are predicted to reach luminosities of $L_\mathrm{max}=(2.4-29.5)\times10^{38}\erg$, or $\log L/\Lsun=4.8-5.9$, which applies to both supernovae that have periodic CCI or just one (Fig.\,\ref{fig:Lmax}). These luminosities are directly correlated to the mass of the companions (Eq.\,\ref{eq:Lmax}) and represent an increase of a factor 3-100 with respect to the companions' pre-supernova luminosities. The distribution of luminosities in supernovae with CCI is instead not distinct from the general population of companions right after the explosion. The only relevant difference is the lack of high-luminosity companions undergoing CCI (Fig.\,\ref{fig:Lmax}), as they form in very massive binaries where the severe mass-loss from both stars substantially widened the orbit before the supernova explosion.

About $7\%$ of H-poor supernovae with periodic CCI (and $5\%$ of all H-poor supernovae more generally) leave a companion whose luminosity after ECI exceeds the \cite{HumphreyDavidson1979} limit, classifying it as a Luminous Blue Variable (LBV). These $\simgr30\Msun$ stars usually remain in this LBV-like state for less than $100\days$, though in some cases the radial expansion persists for $\sim30\yr$. While this mechanism cannot account for the overall LBV population, it makes a testable prediction: late-time observations of faded supernovae may occasionally reveal a surviving LBV-like companion.
 
\begin{figure}
 \centering
    \includegraphics[width=0.94 \linewidth]{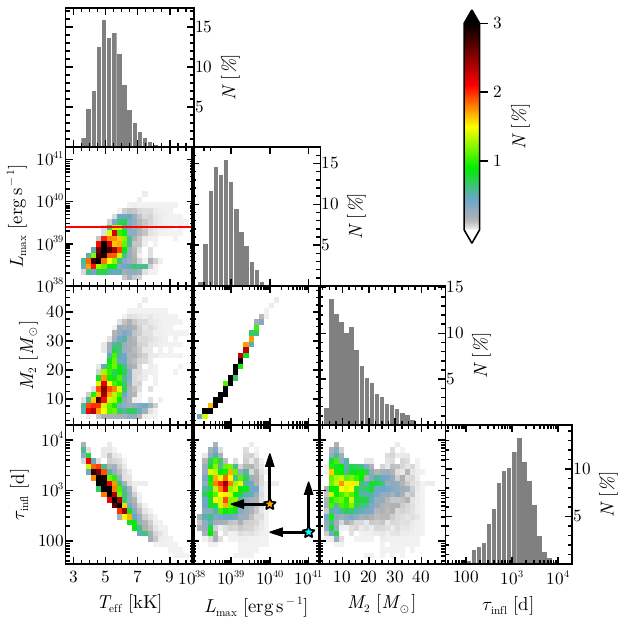}
\caption{As Fig.\,\ref{fig:cornerplot_PeN}, but showing the companion star's mass $M_2$, the timescale for which it will be inflated $\tauinfl$, and its luminosity $L_\mathrm{max}$ and effective temperature $T_\mathrm{eff}$ while inflated. The red line in the $L_\mathrm{max}-T_\mathrm{eff}$ corner-plot is the \cite{HumphreyDavidson1979} limit.}
\label{fig:cornerplot_MLTt}
\end{figure}

Figure\,\ref{fig:cornerplot_MLTt} shows 2D plots of the properties of companion stars of H-poor supernovae that undergo periodic CCI. We find that companion stars generally have typical companion masses $M_2=5-30\Msun$, which are directly linked to their luminosity (Eq.\,\ref{eq:Lmax}): lower-mass companions are dimmer and generally cooler, with cooler companions typically remaining inflated for longer than hotter ones. The companions of systems undergoing \Case BB RLOF have lower masses ($\simle12\Msun$, see Appendix\,\ref{app:parameter_space}) and are therefore much dimmer ($L_\mathrm{max}=2-7\times10^{38}\ergs$).  The companions remain typically inflated for $100-10^4\days$. Within the correlation plot of the inflated luminosity against the inflation timescale (the $L_\mathrm{max}-\tauinfl$ panel), we find that the models cumulate around $L_\mathrm{max}\sim10^{39}\ergs$ (or $\log L_\mathrm{max}/\Lsun=5.4$, which corresponds to $M_2\sim10\Msun$) and $\tauinfl\sim 10^3\days$, so the companions would appear quite bright for a few years post-explosion. 

The distribution of the companion masses $M_2$ roughly follows the IMF, with cutoffs at low ($\simle5\Msun$) and high ($\simgr 40\Msun$) masses. Low-mass companions are missing because low-mass and low mass-ratio binary models often undergo unstable RLOF \citep[see][]{Jin25_BonnGal}, and merge. At high masses, primaries more often produce failed supernovae, and strong wind mass loss widens their orbits before core collapse, making CCI unlikely even if the primary explodes.

\subsection{Different population models}\label{sec:results:multiple}
Several assumptions in our theoretical models remain poorly constrained. In our setup, we can explore variations in (i) the natal kick distribution, (ii) the explodability criteria of stellar models at core collapse (i.e., whether they undergo a successful supernova explosion or instead directly collapse to a BH), and (iii) the criteria for merging a binary during RLOF. 

We define a grid of population models labeled \texttt{k$i$m$j$e$n$}, where $i=1,2,3$ selects the kick prescription, $j=1,\ldots,5$ the merger criteria during the first phase of RLOF, and $n=1,2,3$ the explodability criterion (Table~\ref{tab:acronyms}). The reference model is labeled \texttt{k1m1e1}.
In Sect.\,\ref{sec:results:multiple:kick}, we quantify the effect of adopting different kick distributions, while in Sect.\,\ref{sec:results:multiple:merg_exp} we investigate the impact of modifying the explodability and merger prescriptions.
 
\begin{table}[]
\caption{Parameters explored in different population models. }\label{tab:acronyms}
\resizebox{\columnwidth}{!}{
\begin{tabular}{lll}
Kick         &Pre-supernova merger   &Explodability  \\
distribution &criteria  (first RLOF)    &criteria   \\\hline\hline
{[\texttt{k1}]} \Valli*  & [\texttt{m1}] \citet[] {Jin25_BonnGal}*$^{,\dagger}$           &[\texttt{e1}] \citet[]{MHLC16}* \\
{[\texttt{k2}]} \COMBINE & [\texttt{m2}] \citet[]{Ercolino_widebinary_RSG}                &[\texttt{e2}] \citet[]{Ertl16_explodability_m4mu4, Ertl2020_explodability_HeS} \\
{[\texttt{k3}]} \DM      & [\texttt{m3}] \citet[]{Pavlovskii_Ivanova_2015_MT_from_Giants} &  [\texttt{e3}] \citet[]{PS20_explodability_Mco_Xc} \\
                       & [\texttt{m4}] \citet[]{Pauli_MasterThesis}                     & \\
                       & [\texttt{m5}] \citet[]{PABLO_LMC_GRID}                         & \\
\end{tabular}
}
\tablefoot{Entries are ordered by, in decreasing sequence, the number of binary systems that remain gravitationally bound after the first supernova explosion (Column 1), the number of binaries that do not merge (2), and the number of evolutionary models that successfully explode (3). Each specific population model is identified by the set of adopted parameter labels, which are listed in square brackets in Columns 1, 2, and 3. (*) These correspond to the parameters of the reference model, i.e. model \texttt{k1m1e1}. ($\dagger$) In \PI, this is denoted as the hardcoded merging criterion.  
}
\end{table}

\subsubsection{Different kick distributions}\label{sec:results:multiple:kick}

\begin{figure}
    \centering
    \includegraphics[width=\linewidth]{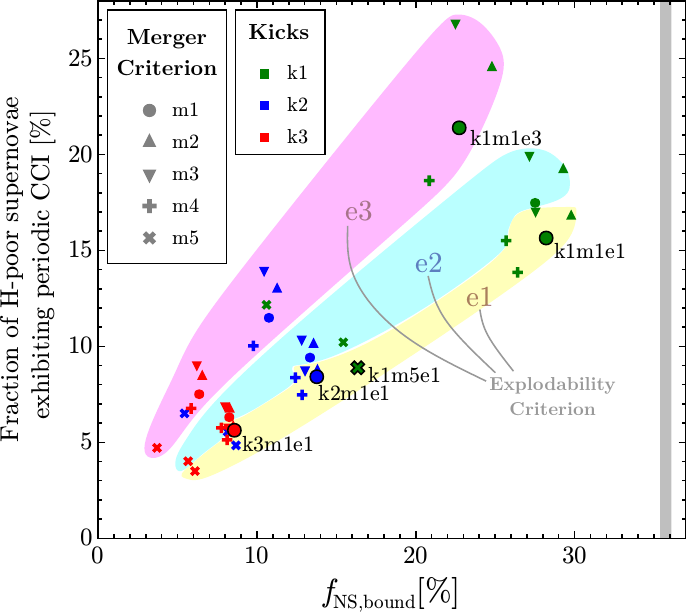}
    \caption{Fraction of H-poor supernovae with periodic companion–compact-object interaction (CCI) for different population models, versus the fraction of binaries in which the primary star explodes and forms a NS that is bound to the companion, $f_\mathrm{NS,bound}$, normalized to all systems where the primary undergoes core collapse. Colors indicate the kick prescription (\texttt{k1}–\texttt{k3}), markers the merger criterion (\texttt{m1}–\texttt{m5}), and shaded regions the explodability criterion (\texttt{e1}–\texttt{e3}; see Table\,\ref{tab:acronyms}) adopted. The five models discussed in the main text are highlighted and labeled. The gray vertical line shows  $f_\mathrm{NS,bound}$ from \citet[see Appendix\,\ref{sec:appendix:OB_cal}]{Schuermann25_SMC_pop_combine}.
    }
    \label{fig:line}
\end{figure}

\EDIT{In this study, the kick distributions can be ranked according to their mean kick magnitude: \Valli\ exhibits the lowest values, followed by \COMBINE, whereas \DM\ yields the highest. For population models differing only in the kick distribution adopted, the fraction of H-poor supernovae showing periodic CCI scales increases with weaker kicks magnitudes, as these leave more binaries bound after explosion (Fig.\,\ref{fig:line}), and viceversa.}

\begin{figure*}
    \centering
    \includegraphics[width=\linewidth]{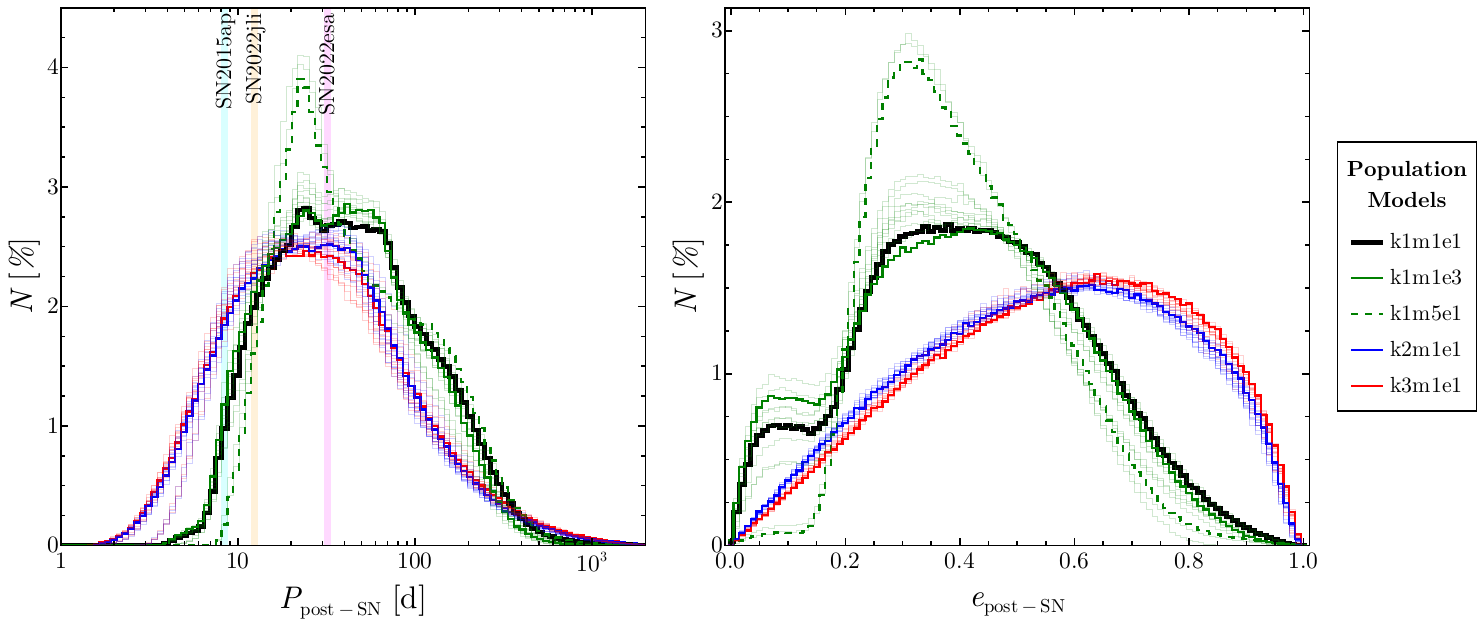}
    \caption{Post-supernova orbital period (left) and eccentricity (right) distributions for binary models where periodic CCI is expected. Each line shows the results from a different population model, with those discussed in Sect.\,\ref{sec:results} shown with thicker lines (see legend) while those with other parameter combinations (see Table\,\ref{tab:acronyms}) are shown with a lighter color, corresponding to the kick distribution adopted (red for \DM, blue for \COMBINE, and green for \Valli). Each histogram is normalized to 100\%. The observed modulation periods of SN\,2022jli \citep{Moore2023_SN2022jli_Ic_Binary,Chen2024_SN2022jli_binary}, SN\,2015ap \citep{Ragosta25_2015ap_CCI}, and SN\,2022esa \citep{Maeda25_2022esa_CSI_or_CCI} are also shown.}
    \label{fig:distr}
\end{figure*}

In model \texttt{k3m1e1}, which \EDIT{only differs from the reference model by employing the kicks of \DM\ instead of \Valli}, the fraction of H-poor supernovae that undergo periodic CCI drops to $6\%$ (or $3-9\%$ across models with the same kick distribution but different explodability and merger criteria; Fig.\,\ref{fig:line}). In contrast, the number of supernovae with only one CCI increases dramatically to about $23\%$ ($14-31\%$). The reduction of the number of supernovae with periodic CCI occurs even though the birth parameter space of their progenitor binaries broadens (cf. Figs.\,\ref{fig:logPq},\ref{fig:logPq_diffpop_1}, and \ref{fig:logPq_diffpop_2}). Systems that remain bound and undergo periodic CCI have, on average, larger post-explosion eccentricities and slightly shorter orbital periods than in the reference population (Fig.\,\ref{fig:distr}).

Model \texttt{k2m1e1} \EDIT{adopts the kicks from \COMBINE, and here the fraction of H-poor supernovae with CCI (both single and periodic) is found between that of the reference model and model \texttt{k3m1e1}. Periodic CCI occurs in} $8\%$ (or $5-14\%$ across models using the same kick distribution but different explodability and merger criteria) of H-poor supernovae, while single CCI in $20\%$ ($13-27\%$).  

\begin{figure*}
    \centering
\begin{tikzpicture}
\node[anchor=south west, inner sep=0] (img){\includegraphics[width=0.497\linewidth]{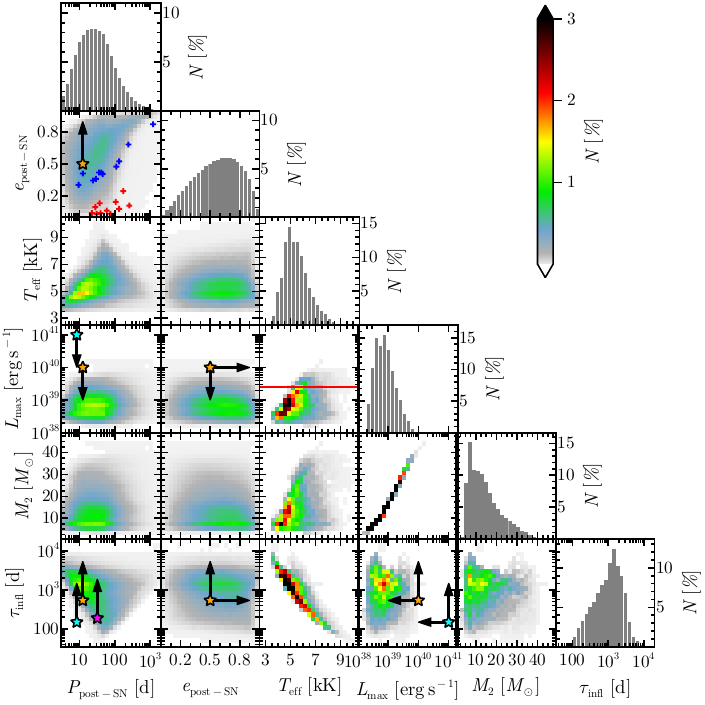}};
\begin{scope}[x={(img.south east)}, y={(img.north west)}]
\node[anchor=north west] at (0.37,.96) {\textbf{Model \texttt{k3m1e1}}};
\end{scope}
\end{tikzpicture}
\begin{tikzpicture}
\node[anchor=south west, inner sep=0] (img){\includegraphics[width=0.497\linewidth]{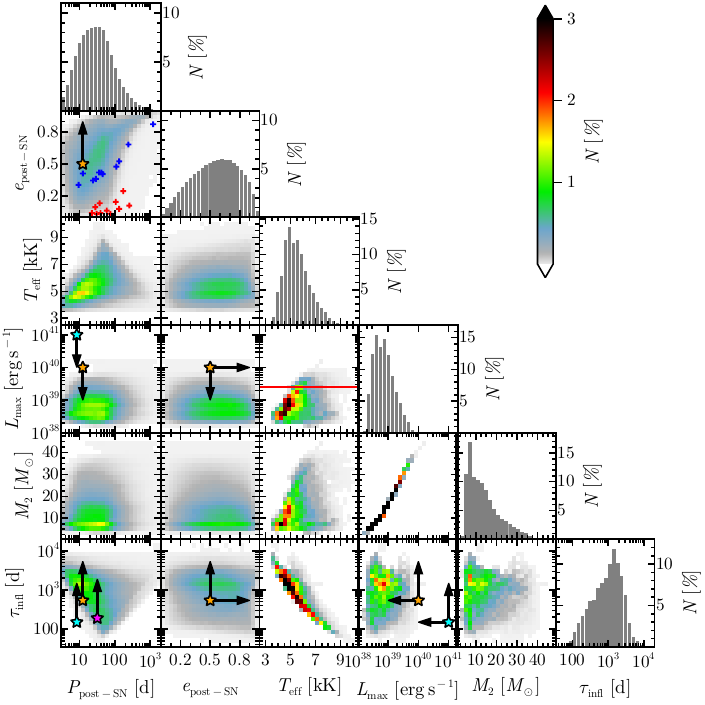}};
\begin{scope}[x={(img.south east)}, y={(img.north west)}]
\node[anchor=north west] at (0.37,.96) {\textbf{Model \texttt{k2m1e1}}};
\end{scope}
\end{tikzpicture}
\begin{tikzpicture}
\node[anchor=south west, inner sep=0] (img){\includegraphics[width=0.497\linewidth]{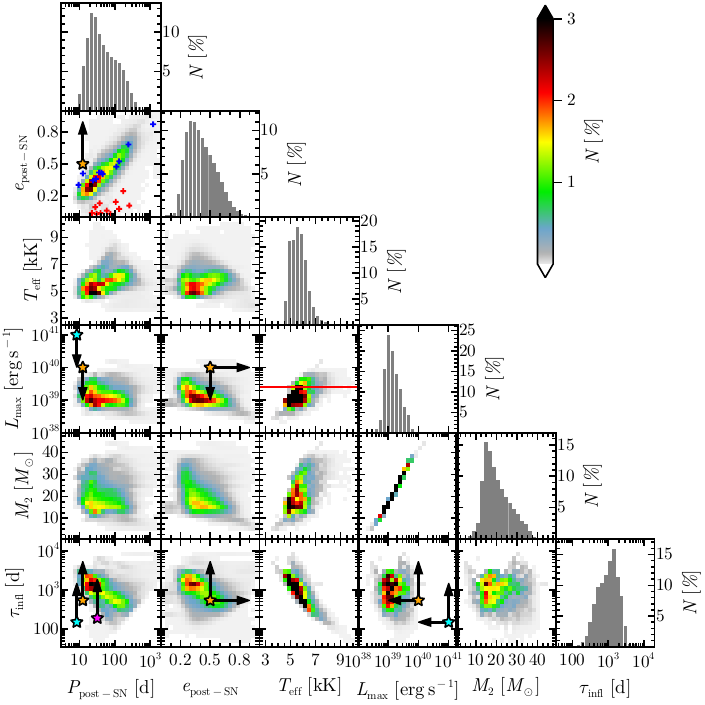}};
\begin{scope}[x={(img.south east)}, y={(img.north west)}]
\node[anchor=north west] at (0.37,.96) {\textbf{Model \texttt{k1m5e1}}};
\end{scope}
\end{tikzpicture}
\begin{tikzpicture}
\node[anchor=south west, inner sep=0] (img){\includegraphics[width=0.497\linewidth]{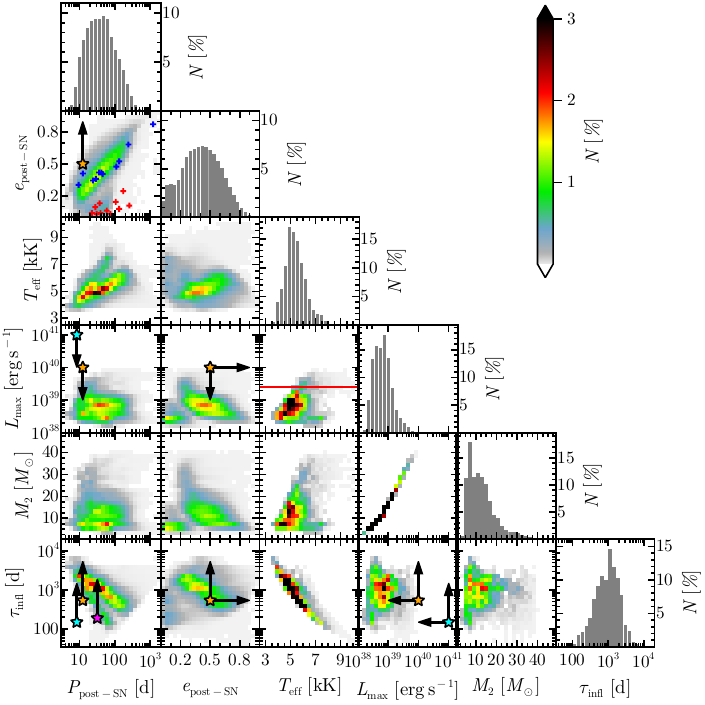}};
\begin{scope}[x={(img.south east)}, y={(img.north west)}]
\node[anchor=north west] at (0.37,.96) {\textbf{Model \texttt{k1m1e3}}};
\end{scope}
\end{tikzpicture}
\caption{Extended cornerplots from Figs.\,\ref{fig:cornerplot_PeN} and \ref{fig:cornerplot_MLTt} using different population models, identified by the label on top. }
    \label{fig:cornerplot_pops}
\end{figure*}

The distributions of post-explosion orbital periods $P_\mathrm{post-SN}$ and eccentricities $e_\mathrm{post-SN}$  in binaries whose first supernova undergoes periodic CCI show no significant differences between the \COMBINE\ and \DM\ kick prescriptions (Fig.\,\ref{fig:distr}). In model \texttt{k2m1e1}, a notable feature is the absence of a clearly defined low-eccentricity subpopulation. This is unexpected, since the kick distribution includes a low-kick component analogous to that in \Valli, where such a subpopulation is clearly visible (Fig.\,\ref{fig:distr}). The apparent absence of this component with the \COMBINE\ kicks instead arises from systematically stronger kick magnitudes in the low-kick group, \EDIT{resulting in a broader eccentricity distribution for this group as well, thus smearing their contribution}. As a result, most of the structure in the $P_\mathrm{post-SN}$–$e_\mathrm{post-SN}$ corner-plot in Fig.\,\ref{fig:cornerplot_PeN} vanishes for both models \texttt{k2m1e1} and \texttt{k3m1e1} (Fig.\,\ref{fig:cornerplot_pops}).

\EDIT{Generally, all population models implementing the kicks from \Valli\ yield qualitatively different eccentricity distributions compared to models that implement the \COMBINE\ or \DM\ kicks. This also results in systematically larger orbital periods, since the polar-aligned kicks in \Valli\ } can only widen the orbit. Nonetheless, the orbital period  distribution in binaries which undergo periodic CCI following the explosion of the primary are broadly similar \EDIT{regardless of the kick distribution adopted, as all population models} share a peak around $20-50\days$ (Fig.\,\ref{fig:distr}). Similarly, the properties of the companions in these binaries are not significantly different between models \texttt{k1m1e1}, \texttt{k2m1e1} and \texttt{k3m1e1} (Figs.\,\ref{fig:cornerplot_MLTt}, \ref{fig:cornerplot_pops}).

\subsubsection{Different merger and explodability criteria}\label{sec:results:multiple:merg_exp}

The reference population model produces the fewest binaries that (i) merge after the first phase of RLOF turns unstable and (ii) form BHs following core collapse. It therefore yields the largest birth-parameter space for systems undergoing CCI, compared to models with different merger and explodability prescriptions (see Table\,\ref{tab:acronyms} and \PI).

\begin{figure}
    \centering
    \includegraphics[width=\linewidth]{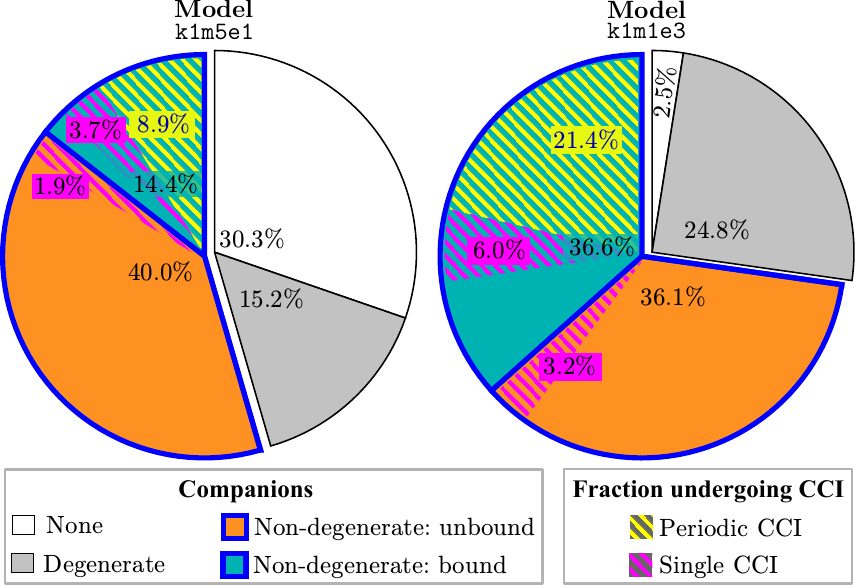}
    \caption{As Fig.\,\ref{fig:pies}, but for population models producing more pre-supernova mergers (\texttt{k1m5e1}, left) or less supernovae (\texttt{k1m1e3}, right).}
    \label{fig:pies_multi}
\end{figure}

\EDIT{To explore the effect that different merger criteria have on the population of H-poor supernovae undergoing CCI, we investigate model \texttt{k1m5e1}, which adopts the merger criteria resulting in the largest fraction of binary evolution models merging. Here, the binaries that merge typically have lower initial primary masses, small initial orbital separations, and low companion masses}. In the reference model, these systems were favored to produce supernovae with periodic CCI (cf. Fig.\,\ref{fig:logPq} and Fig.\,\ref{fig:logPq_diffpop_3}). Compared to the reference model, the number of H-poor supernovae whose progenitor had a binary companion decreases  ($54\%$), even more \EDIT{so} for those where the binary remains bound after the explosion ($14\%$, Fig.\,\ref{fig:pies_multi}). As a result, model \texttt{k1m5e1} produces few H-poor supernovae exhibiting periodic CCI ($\sim 9\%$ of the total; see Fig.\,\ref{fig:pies_multi}). This smaller number of supernovae also feature overall more massive companions than in the reference model (as those with $\simle15\Msun$ have instead merged), which are typically brighter and colder with respect to the population from the reference model (cf. Fig.\,\ref{fig:cornerplot_pops}). 
The orbital period distribution is qualitatively affected, by removing the short period systems ($\simle 8\days$) and reducing the contribution of intermediate periods ($30-200\days$, see Fig.\,\ref{fig:distr}), compared to the reference model. The eccentricity distribution is also affected, by completely removing the low-end of the eccentricity distribution (Fig.\,\ref{fig:distr}), since it is contributed predominantly by systems which underwent \Case BB RLOF (see Sect.\,\ref{sec:example:CCI_properties} and Appendix\,\ref{app:parameter_space}). \EDIT{Population models that adopt the other merging criteria listed in Table\,\ref{tab:acronyms} result in intermediate amounts of binary evolution models merging. Model \texttt{k1m4e1} affects a similar parameter space as \texttt{k1m5e1}}, but the magnitude of its effects is less significant. The opposite is true in models \texttt{k1m3e1} and \texttt{k1m2e1}, where the merger criteria preferentially merge wider binaries, which are not expected to give rise to supernovae exhibiting CCI. \EDIT{Here}, the number of supernovae exhibiting CCI increases with respect to the reference population model, although not significantly (Fig.\,\ref{fig:line}). Model \texttt{k3m5e1}, in particular, produces the least number of H-poor supernovae with periodic CCI across all population models ($3\%$).

\EDIT{We then explore the effect that explodability criteria producing less supernovae would have on the population of H-poor supernovae undergoing CCI. In model \texttt{k1m1e3}, the different explodability criterion results in a substantial fraction of the high-mass progenitors of \Type Ibc supernovae imploding rather than exploding. This mainly concerns \Type Ibc progenitors coming from merger products, single stars, and secondaries, while it does not affect stars stripped from binary interactions significantly (\PI). As a result, the population of would-be H-poor supernovae with no companions almost completely vanishes ($2.5\%$), which boosts the fraction of those with a degenerate or non-degenerate companion ($24.8\%$ and $72.7\%$ respectively, see Fig.\,\ref{fig:pies_multi})}.
The relative fraction of H-poor supernovae with periodic CCI therefore increases to $21\%$ (Fig.\,\ref{fig:pies_multi}). In terms of companion properties in these supernovae, the contribution from larger mass companions ($\simgr20\Msun$) is diminished but not removed entirely, and there are no significant changes in terms of luminosity and temperature distribution of the companions (cf. Fig.\,\ref{fig:cornerplot_pops}). Similarly, the orbital period and eccentricity distributions in this population model are roughly similar to those of the reference population model (Fig.\,\ref{fig:distr}). We find that model \texttt{k1m3e3} in particular predicts the largest fraction of H-poor supernovae with periodic CCI ($27\%$ of all H-poor supernovae). 
Model \texttt{k1m1e2} shows the same differences from the reference model as \texttt{k1m1e3}, but less pronounced.

\section{Comparison to observed transients}\label{sec:observations}
To date, only two core-collapse supernovae have been reported with periodic light curve modulations coinciding with H$\alpha$ velocity shifts: \SN{2022jli} \citep{Moore2023_SN2022jli_Ic_Binary,Chen2024_SN2022jli_binary} and \SN{2015ap} \citep{Ragosta25_2015ap_CCI}. In the following, we investigate whether the available observational constraints for \SN{2022jli} (Sect.\,\ref{sec:observations:2022jli}) and \SN{2015ap} (Sect.\,\ref{sec:observations:sn2015ap}) can be explained with our stellar evolution models. Furthermore, we provide a brief discussion of \SN{2022esa} (Sect.\,\ref{sec:observations:sn2022esa}), as well as of other supernovae for which only photometric modulations have been documented (Sect.\,\ref{sec:observations:other_transients}).

\subsection{SN2022jli}\label{sec:observations:2022jli}
\cite{Moore2023_SN2022jli_Ic_Binary} and \cite{Chen2024_SN2022jli_binary} report a post-explosion flux modulation with a period of $12.4\pm0.1\days$, which emerged approximately $50\days$ after the first detection and persisted until the last detection in the optical. These modulations were also detected in the $\gamma$-ray flux for $1.5\yr$ following the explosion \citep{Zhang25_2022jli_gammarays}. \EDIT{Assuming that periodic CCI produces the modulation in the optical and $\gamma$-ray fluxes through (periodic) accretion onto the compact object, we can estimate} a lower bound on $\tauinfl$ and $N_\mathrm{inter}$ (see Sects.\,\ref{sec:method:ECI} and \ref{sec:method:CCI}).
In \cite{Chen2024_SN2022jli_binary}, the most recent detection of the supernova yielded a bolometric luminosity of approximately $\sim 10^{40}\ergs$, which provides an upper bound on the companion’s maximum luminosity, $L_\mathrm{max}$.

To identify binary evolution models capable of reproducing the observed properties of \SN{2022jli}, we impose two additional selection criteria.  
First, we require that the post-explosion binary orbit has a high eccentricity ($e_\mathrm{post-SN}>0.5$), as motivated by \citet{Chen2024_SN2022jli_binary} and \citet{Hirai25_2022jli_ECI}. 
Second, we require that the ejecta be helium-poor \citep[where we consider a threshold for the pre-explosion surface helium abundance of the progenitors $Y_\mathrm{pre-SN}<0.5$;  see;][]{Dessart2020_Ibc}, so as to emulate the \Type Ic spectroscopic classification of the progenitor.

Within the reference population model, two binary evolution models satisfy the constraints inferred for \SN{2022jli} (Table\,\ref{tab:inferred_data_2022jli}). The ejecta mass in these models is $3.8-3.9\Msun$, with little He ($0.16-0.17\Msun$), consistent with the \Type Ic classification (although the He mass alone may not distinguish \Type Ib from \Type Ic supernovae, see \citealt{Yoon15_Ibc_Hecontent, Dessart2012_IbcSNe, Dessart2020_Ibc}).  The inferred eccentricity, while high ($0.52-0.57$), is still lower than those reported in the exploratory models of \citet[$e_\mathrm{post-SN}\simgr0.7$]{Chen2024_SN2022jli_binary} and \citet[$0.7\simle e_\mathrm{post-SN}\simle0.8$]{Hirai25_2022jli_ECI}.  ECI-driven inflation lasts up to $12\yr$ (but this should be taken as an upper-limit, see Sect.\,\ref{sec:disc:ECI_CCI:CCI}), during which the companion achieves and maintains a luminosity of $L_\mathrm{max}=3.5\times10^{38}\ergs$, about 30 times fainter than the last detection in \citet{Chen2024_SN2022jli_binary}. We use this luminosity to determine the companion star's magnitude with the VLT's HAWK-I instrument (Appendix\,\ref{sec:appendix:observability}). We find that the companion would have a $J$-band magnitude of $\sim22.6$. Pre-supernova observations of the local environment with only a 4-minute exposure time could detect sources down to a $J$-band magnitude of $\sim23$ \citep{Grosbol12_straformation_GDgalaxies}. Combined with the long timescale inferred for its inflation, we expect the companion will be observable today.

\begin{table}
    \centering
    \caption{Observational constraints (top; see text for references) and inferred properties (bottom) of \SN{2022jli}, \SN{2015ap}, and \SN{2022esa} from the reference population model. An extended version with inferred parameters from other population models is available in Table\,\ref{tab:2022jli_ext}.}
\resizebox{.95\columnwidth}{!}{
    \begin{tabular}{r||ccc}
    Observational \\
     constraints   & \SN{2022jli} & \SN{2015ap} & \SN{2022esa}
     \\\hline\hline
     $P_\mathrm{post-SN} \ [\days]$  &
     $12.4\pm0.1$ & $8.41\pm0.08$ & $28.8-34.6$\\[3pt]
     $\tauinfl \ [\days]$  &
     $\geq540$ & $\geq175$ & $\geq190$ \\[3pt]
     $e_\mathrm{post-SN}$  & 
     $>0.5$ & $\times$ & $\times$\\[3pt]
     $Y_\mathrm{pre-SN}$  &
     $<0.5$ & $>0.5$ & $<0.5$\\[3pt]
     $M_\mathrm{ej} \ [\Msun]$ & 
     $\times$ & $2.2\pm0.6$ & $\times$  \\[3pt]
     $L_\mathrm{max} \ [10^{38}\ergs]$ & 
     $\simle 100$ & $\simle 1000$ & $\times$  \\[3pt]
     \multicolumn{1}{l}{ } & 
\multicolumn{1}{c}{ } & 
\multicolumn{1}{c}{ } \\
     Inferred  \\
     parameters  & \SN{2022jli} & \SN{2015ap} & \SN{2022esa}
     \\\hline\hline
    $M_\mathrm{1,i} \ [\Msun]$ &
   $28.2$ &	$15.9-31.6$ &	$28.2-44.7$\\[3pt]
     $M_\mathrm{2,i} \ [\Msun]$ & 
     $7.1$ & $7.1-15.1$ & $7.1-33.5$ \\[3pt]
     $\qi=M_\mathrm{2,i}/M_\mathrm{1,i}$ & 
     $0.25$ & $0.35-0.60$ & $0.25-0.85$ \\[3pt]
     $P_\mathrm{i} \ [\days]$ &
     $25.1- 28.2$ & $2.8-10.0$ & $5.6-31.6$\\     
     Binary models & 2 & 77 & 230\\[3pt]\hline
    $M_\mathrm{ej} \ [\Msun]$ & 
     $3.8-3.9$ &	$1.8-2.8$ & $3.8-5.1$ \\[3pt]
     $M_\mathrm{He,ej} \ [\Msun]$  &
     $0.17-0.18$ & $1.2$ & $0.17-0.27$\\[3pt]
          $Y_\mathrm{pre-SN}$  &
     $0.30-0.32$ & $0.98$ & $0.21-0.48$ \\[3pt] \hline
     $M_2 \ [\Msun]$ &  
     $7.4$ & $7.4-17.8$ & $7.3-34.5$  \\[3pt]
      $\tauinfl \ [\yr]$ &
      $10.8-11.8$ & $4.1-16.3$ & $1.3-9.9$ \\[3pt]
       $L_\mathrm{max} \ [10^{38}\ergs]$ &
      $3.5$  & $ 3.5-11.2$ & $3.5-45.0$\\[3pt]
     $T_\mathrm{eff} \ [\mathrm{kK}]$ &
   $4.0$ & $3.8-4.8$ & $4.1-6.5$ \\
$e_\mathrm{post-SN}$  & 
     $0.52-0.57$ & $0.18-0.40$ & $0.25-0.64$\\[3pt]
    \end{tabular}
}
    \tablefoot{Observational constraints include the post-supernova modulation period ($P_\mathrm{post-SN}$), the duration over which the companion star remains inflated ($\tauinfl$; see text), the post-supernova orbital eccentricity ($e_\mathrm{post-SN}$), the \EDIT{surface} helium mass fraction inferred from the supernova spectral classification ($Y_\mathrm{pre-SN}$), the ejecta mass ($M_\mathrm{ej}$), and upper limits on the luminosity of the companion star ($L_\mathrm{max}$). 
    The inferred quantities comprise three main categories. The upper part shows the ZAMS progenitor binary configuration, including the component masses ($M_\mathrm{1,i}$, $M_\mathrm{2,i}$), mass ratio ($\qi$), and orbital period ($P_\mathrm{i}$), as well as the number of distinct binary evolution models that are consistent with the applied constraints. 
    The middle part shows the properties of the supernova progenitor, including the ejecta mass ($\Mej$), the helium mass contained in the ejecta ($M_\mathrm{ej,He}$), and the progenitor’s surface helium mass fraction prior to core collapse ($Y_\mathrm{1,pre-SN}$). The lower part shows the post-supernova properties of the companion star following ECI (its mass $M_2$, its inflation timescale $\tau_\mathrm{infl}$, and its maximum luminosity $L_\mathrm{max}$ and effective temperature $T_\mathrm{eff}$ during the inflated phase) as well as the post-supernova orbital eccentricity ($e_\mathrm{post-SN}$).
}
    \label{tab:inferred_data_2022jli}
\end{table}

We can also infer the properties of the progenitor binary system at birth: the ZAMS mass of the supernova progenitor is $M_\mathrm{1,i}=28.2\Msun$, that of the companion is $M_\mathrm{2,i}=7.1\Msun$, and the initial orbital period is $25.1-28.2\days$. The corresponding evolutionary models undergo \Case B RLOF, which means that the NS received a polar kick (Sect.\,\ref{sec:method:kicks}). This, combined with the narrow kick magnitude distribution (Table\,\ref{tab:kicks}) strongly limits the initial parameter space possible to produce a $12.4\days$ binary after explosion, as it requires a relatively fine-tuned ejecta mass and lower pre-explosion orbital period. Additionally, these binary evolution models are close to the part of the parameter space where \Case B RLOF turns unstable (specifically for lower $\qi$, $P_\mathrm{i}$, and $M_\mathrm{1,i}$), which further narrows down the parameter space of progenitor models.

When analyzing different population models, particularly those employing the \COMBINE\ and \DM\ kicks, the birth parameter space producing events compatible with \SN{2022jli} expands to encompass higher primary and companion masses. The initial orbital periods are also broader, but always encompassing binaries that have undergone RLOF prior to the supernova. Since the companion can be more massive,  it may be easier to detect in follow-up observations (see Appendix\,\ref{sec:appendix:observability}). At the same time, the inflation timescale $\tauinfl$ can be significantly shorter than in the reference model, implying that the companion star may already have re-established thermal equilibrium. Finally, the inferred post-supernova eccentricities $e_\mathrm{post-SN}$ can reach values as high as $\sim0.8$, in closer agreement with the results of \cite{Chen2024_SN2022jli_binary} and \cite{Hirai25_2022jli_ECI}. Only a few population models fail to identify any binary model consistent with the observational constraints, specifically \texttt{k1m5e1}, \texttt{k1m5e2}, and \texttt{k1m5e3}, for which all candidate binaries identified in \texttt{k1m1e1} are instead expected to merge before the first explosion.
 
 It is worth noting that future observations, if taken before the year 2035, can update the constraints on the companion's luminosity $L_\mathrm{max}$ and its inflation timescale $\tauinfl$, even with a non-detection. We also provide predictions on the companion's magnitude for future detections with HST and JWST (see Table\,\ref{tab:2022jli_HST_JWST}). Obtaining stricter constraints would narrow down the population models that would replicate \SN{2022jli}, as well as the birth parameter space of the binary progenitor. 

\subsection{SN2015ap}\label{sec:observations:sn2015ap}
\cite{Ragosta25_2015ap_CCI} reported an undulation period of $8.41\pm0.08\days$ in the light curve of the Type Ib supernova \SN{2015ap}, which is consistent with oscillatory behavior observed in the H$\alpha$ line, suggesting that \SN{2015ap} is analogous in nature to \SN{2022jli}. From the duration of the light curve, we infer an inflation timescale of $\tauinfl>175\days$. Given that this transient is a \Type Ib supernova, we adopt a surface helium mass fraction opposite to that inferred for the progenitor of \SN{2022jli}, namely $Y_\mathrm{pre-SN}>0.5$ (see Sect.\,\ref{sec:observations:2022jli}). We estimate an upper limit on the peak luminosity of $L_\mathrm{max}<10^{41}\ergs$ based on the supernova's luminosity at the last reported detection, and use the the ejecta mass $\Mej$ estimates by \cite{Ragosta25_2015ap_CCI} as an additional constrain.

Within the reference population model, we identify approximately 70 distinct binary evolution models that reproduce the observed properties of \SN{2015ap}. These models all share $Y_\mathrm{pre-SN}=0.98$, consistent with the \Type Ib classification. The luminosity $L_\mathrm{max}$ of the inflated companion is 90-300 times lower than the last detection of the supernova, and follow-up Hubble Space Telescope observations obtained in 2016, 2017 and 2020 do not reveal any luminous source at the supernova position \citep{Aryan21_sn2015ap_sn2016bau_progenitor}. The inferred inflation timescale of the companion $\tauinfl$ likewise indicates that the companion star may already have returned to thermal equilibrium.

All our population models are capable of finding binary evolution models capable of producing \SN{2015ap}. Like for \SN{2022jli}, all these binary models have undergone RLOF during their evolution before the supernova. When dissecting the results of different population models, the most significant difference lies with the eccentricity distributions, which can be significant ($e_\mathrm{post-SN}=0.8$) only when adopting the kick distributions of \COMBINE\ or \DM.

\subsection{SN2022esa}\label{sec:observations:sn2022esa}
The \Type Ic supernova \SN{2022esa} has a $\sim 29$–$35\days$ periodic signal in its light curve \citep{Maeda25_2022esa_CSI_or_CCI}, which remains detectable over the observed duration of the event ($\sim190\days$). The transient does not show a variable H$\alpha$ emission line, which may suggest that the energy source responsible for the observed luminosity undulations is not necessarily CCI \citep{Maeda25_2022esa_CSI_or_CCI}.

Nonetheless, we investigate our synthetic binary populations to identify systems consistent with the inferred properties of \SN{2022esa}, since our models can produce \Type Ic supernova progenitors (Sect.\,\ref{sec:observations:2022jli}). Owing to the few observational constraints, we recover a substantial number of binary evolution models ($\sim200$) that reproduce the features of \SN{2022esa}. Similarly to the cases of \SN{2022jli} and \SN{2015ap}, all of these progenitor systems experience RLOF before the explosion.   

    \EDIT{\cite{Maeda25_2022esa_CSI_or_CCI} suggests that the undulation period in \SN{2022esa} is increasing, though this does not appear to be statistically significant. This trend contradicts the anticipated behavior for ongoing CCI, where drag on the NS should shorten the orbital and undulation periods. One process that could explain this is the interaction with shellular circumstellar material \citep{Maeda25_2022esa_CSI_or_CCI}.} 

\subsection{Other transients}\label{sec:observations:other_transients}
In principle, single CCI could account for the occurrence of a second peak in double-peaked supernovae\EDIT{, especially if the interaction is delayed. This is possible for a wide pre-explosion binary that reaches periastron at a later time following the explosion. In our reference population model, we find that a single phase of CCI due to one periastron passage occurring during the companion's inflated phase} happens in $7.7\%$ of H-poor supernovae (see Sect.\,\ref{sec:example:CCI_numbers}). If this interaction contributes a sufficiently large luminosity, it may also explain double-peaked supernovae referred to as ``bactrian'' supernovae,  where the extra peak can be as bright, if not brighter than the main one (e.g., \SN{2022xxf}, \citealt{Kuncarayakti2023_SN2022xxf_IcBL_multipeak} and \SN{2020acct}, \citealt{Angus24_2024acct_doublepeaked}). This underscores the potential observational importance of CCI in shaping supernova light curve morphology. \EDIT{It is worth noting that these supernovae with a strong second peak in the light curve could be explained following the merger of the compact object with the companion} \citep{HiraiPosdi22_ECI_NS}. 

A fraction of H-poor superluminous supernovae (SLSNe-I) display periodic undulations in their light curves \citep{Chen2023_SLSNEI_CSM_Magnetar_undulations, Hosseinzadeh22_SLSNe_bumps_and_oscillations}, with a mean observed period of $28.8^{+14.4}_{-9.1}\days$ \citep{Chen2023_SLSNEI_CSM_Magnetar_undulations}. There are also studies focusing on individual SLSNe-I with these modulations have been analyzed in detail, like \object{SN2020qlb} \citep{West_2020qlb_SLSNI_oscillations}, \object{SN2015bn} \citep{Nicholl2016_2015bn_SLSN_undulation}, and \object{SN2024afav} \citep{Farah25_SLSNI_undulations_magnetar}.
Although SLSNe-I are not expected to arise from our binary evolution models, as their progenitor stars are expected to form preferentially in environments with metallicities lower than those assumed in our models \citep{Perley16_hostgalaxies_SLSNe,Schulze18_hostgalaxies_SLSNe}, their progenitors may still reside in binary systems. Therefore, the occurrence of CCI in these transients cannot be excluded. 

\EDIT{The H-rich material ablated from the companion during ECI, or stripped through single or periodic CCI, could produce narrow H lines in the spectra once the supernova photosphere recedes sufficiently. This could explain the appearance of H$\alpha$ in the \Type Ib \SN{2019yvr} \citep{Kilpatrick2021_2019yvr_latetimeinteraction,Ferrari2024_Sn2019yvr_latetimeH_interaction, BaerWay2026_Sn2019yvr_lateH_CSM}. There is also a growing number of H-poor superluminous supernovae where a H$\alpha$ line appears at late times  \citep{Yan2017_SLSNeI_withHalpha_and_oscillations,Pursiainen2022_2018bsz_SLSNI_withH}, and \object{iPTF15esb} is particularly suggestive as it also exhibits three bumps \EDITTWO{in the light curve} about $\sim22\days$ apart from each other \citep{Yan2017_SLSNeI_withHalpha_and_oscillations}, perhaps hinting at CCI being at play}.

Some \Type IIn supernovae also have multiple bumps in their light curves \citep{Nyholm2020_IInSNe_bumps}, especially \object{iPTF13z} \citep{Nyholm17_iPTF13z_IIn_with_undulations_bump}, although their periodicity is less clear. These supernovae may be produced in our binaries \citep{Ercolino_widebinary_RSG}, and a subset of these supernovae may potentially develop CCI (see Sect.\,\ref{sec:discussion:additionalpathways:CEE}). Finally, there are events like \SN{2022mop}, where the oscillation is thought to arise via CCI where the compact-object probes deep inside the companion's envelope, ultimately triggering a merger-burst \citep{Brennan2025_SN2022mop}.

The observed undulations in many transients may alternatively be explained through other mechanisms, such as interaction with concentric shells of circumstellar material \citep{ Nyholm17_iPTF13z_IIn_with_undulations_bump,Chen2023_SLSNEI_CSM_Magnetar_undulations,Nyholm2020_IInSNe_bumps,Maeda25_2022esa_CSI_or_CCI} or some central engine \citep{Chen2023_SLSNEI_CSM_Magnetar_undulations} such as newly formed magnetar \citep{Farah25_SLSNI_undulations_magnetar}. In some transients, such as the \Type Ibc supernova \SN{2019tsf} \citep{Sollerman2020_2019tsf_2019oys, Zenati2025_SN2019tsf_three_peaked_Ib}, the multiple ``bumps'' observed in the light curve may be accounted for by interaction with circumstellar shells produced in a triple system.

\section{Discussion}\label{sec:discussion}
The results shown in this work are contingent upon a number of underlying assumptions. We discuss those in the grids of stellar-evolution models on which the population-synthesis code is based (Sect.\,\ref{sec:disc:models}), as well as for the assumptions adopted in the population-synthesis calculations (Sect.\,\ref{sec:discussion:additionalpathways}). We further analyze the impact of the assumptions made in the modeling of ECI and CCI (Sect.\,\ref{sec:disc:ECI_CCI}), and we discuss additional influences that CCI may exert on the properties of the NS, the companion star, and the binary system as a whole (Sect.\,\ref{sec:discussion:futureevo}). Finally, we briefly compare some of our results with those of \cite{Zapartas25_demographics_companions_of_SESNe} in Sect.\,\ref{sec:discussion:compare}.

\subsection{Stellar and binary physics}\label{sec:disc:models}

Our binary evolution models assume rotation-limited accretion during RLOF, typically yielding accretion efficiencies $\simle10\%$. However, recent studies indicate that the accretion efficiency may be higher \citep{Schootemeijer18_PhiPersei, Vinciguerra2020, Xu25_SMC_pop_mesa, Schuermann25_SMC_pop_combine, Lechien25_accretionefficiency}, while still not fully conservative \citep{Zapartas25_demographics_companions_of_SESNe}. Varying the accretion efficiency, and similarly the angular momentum carried away by non-accreted material, changes the pre-explosion orbit and the companion’s mass and radius, which then modify the amount of energy intercepted by the companion  (Eqs.\,\ref{eq:Eheat},\ref{eq:OmegaEff}), and in turn the magnitude and duration of the companion’s swelling following ECI (Eqs.\,\ref{eq:Rmax},\ref{eq:tau_ECI}). These changes also influence whether the binary remains gravitationally bound following the natal kick.
Higher accretion efficiencies can also induce significant radial expansion of the companion during RLOF \citep{SchuermannLanger25_stableRLOF_accretor, Wang26_accretion_and_criticalrotation}, which can promote the onset of unstable RLOF prior to the first supernova in tighter binaries, which would likely reduce the number of systems producing supernovae with CCI in a similar manner to when we applied the merger criterion \texttt{m5} in Sect.\,\ref{sec:results:multiple:merg_exp}.

The predicted number of supernovae with CCI is also sensitive to wind mass loss and metallicity.  The wind mass-loss prescription applied to stripped stars is uncertain, and in our models they are likely overestimated (see Appendix\,A.1.1 in \PI). A reduction of the mass-loss rates in these systems would result in more compact pre-supernova orbits, which consequently increases the predicted number of supernovae exhibiting CCI. Combined with weakened winds from lower metallicities, this could decrease the fraction of H-poor supernovae \citep{Souropanis25_SESNe_Metallicity_POSYDON} and allow some H-deficient supernovae (i.e. \Type IIb) to also exhibit CCI.

\subsection{Population synthesis assumptions}\label{sec:discussion:additionalpathways}

\subsubsection{Post common-envelope systems}\label{sec:discussion:additionalpathways:CEE}
Our population models only consider supernovae originating from binaries that have undergone stable RLOF prior to the first supernova. The evolutionary pathways involving systems in which RLOF has turned unstable are omitted, under the assumption that such systems merge following a common-envelope phase (\PI).

However, a subset of systems that experience unstable \Case C RLOF could potentially eject the common-envelope, according to the $\alpha\lambda$ energy formalism \citep[][assuming $\alpha=1$]{Webbing_alphalambda,deKool_CEE,DewiTauris_CEE,Wang_CEE} or have the primary explode before the merger occurs  \citep{Ercolino_widebinary_RSG}. In both cases, the binary is expected to be very tight at the time of the supernova and it is therefore likely that the supernova would exhibit CCI features. It is worth noting that, due to the close-by circumstellar medium (which can give rise to \Type IIn or even SLSNe-II supernovae, \citealt{Ercolino_HpoorInteractingSNe}), the signatures of CCI would compete with those from circumstellar medium interaction. 

We also find a small number of systems undergoing unstable \Case B RLOF which may survive the common-envelope phase, again according to the $\alpha\lambda$ energy formalism, but they ultimately contribute marginally to the population of supernovae with CCI. Some systems also experience inverse RLOF, as the secondary star transfers mass to the primary before the latter reaches core collapse, which we also assume to become unstable. Even if some binaries were to eject the common-envelope, both stars will evolve as stripped stars, and their small radii will likely prevent the development of CCI.

\subsubsection{Black holes}

In the analysis presented here, we neglect the systems where core collapse of the primary is expected to produce a BH (Sect.\,\ref{sec:methods}) as they receive no kicks and produce no ejecta to trigger ECI. However, this is a oversimplification, as BH formation is predicted to, in some cases, occur in the presence of an energetic explosion \citep{Chan2018_ccsn_explosion_fallback_BH,Chan2020_BH_formation_from_CCSNe_fallback,Burrows25_BH_formation_CCSN}, which is also inferred in some transients \citep{Perley2022_SN2021csp_Icn, Ziimerman2024_2023ixf}. Similarly, BHs may receive natal kicks, as indicated by both theoretical works \citep{Janka2013_kicks_BHs,JankaKresse24_NS_BH_kicks,Burrows24_NS_BH_kicks} and observational evidence \citep{RepettoDaviesSigurdsson2012_SMBH_kicks_similartoNS, Nagarajan_ElBadry24_natalkicks_for_BHs}.
CCI may therefore still occur even when a BH is formed, in which case the higher mass of BHs may remove more of the companion’s inflated envelope through higher accretion rates predicted within the \citet{BondiHoyle44_accretion} scheme. 

\subsubsection{Other phenomena affecting the orbital parameters}
\EDIT{Our models neglect phenomena like the NS rocket mechanism \citep{HarrisonTademaru75_NSRocket, Hirai24_NSRocket}, as it acts on timescales typically (but not always) longer than the post-explosion orbital period in our models undergoing CCI. The secular effect of the rocket mechanism may not affect the orbital period \citep{Hirai24_NSRocket} while it will alter its eccentricity, and therefore the periastron distances. As such, the varying periastron distance may potentially trigger CCI only after some time, or prematurely terminate it if it was already underway.}

\EDIT{ECI can also impart a kick of order $1-100\kms$ to the companion star, increasing in magnitude for tighter binaries and for more energetic explosions \citep{Hirai18_ECI}. This modifies the post-explosion orbital parameters \citep[which are still derivable self-consistently,][]{TaurisTaken98_binarypostExp_parameters_kicks_impact}. Such kick will increase the post-explosion orbital period, whereas the resulting eccentricity depends on both the magnitude and direction of the natal NS kick. For the population of H-poor supernovae undergoing CCI, the expected population-wide effect would be to shift the lower end of the post-explosion period distribution (Fig.\,\ref{fig:distr}) toward longer periods.}

\subsection{The modeling of ECI and CCI}\label{sec:disc:ECI_CCI}
\subsubsection{ECI}
In this work, ECI is modeled using the fitting relations provided by \cite{Ogata21_ECI}. They show that the radii of the companions following ECI can be lower than those yielded by Eq.\,\ref{eq:Rmax}. Consequently, the predicted number of supernovae exhibiting CCI should be taken as an upper limit, and the true distribution of the parameters we presented may be narrower than we inferred here.  

Moreover, the stellar evolution models used in \cite{Ogata21_ECI} to produce the fitting relations differ from those employed in our population synthesis, as they neglect the effects of binary interaction and rotation. In our calculations, approximately two thirds of the companions of supernovae exhibiting CCI are rapidly rotating at the time the primary explodes, with velocities exceeding $80\%$ of their critical value (which is also subject to modeling uncertainties, see \citealt{Wang26_accretion_and_criticalrotation}).
\EDIT{The rotation rate of the companion star is expected to decrease significantly after ECI \citep{Pan2012_ECI_slowdown_companionrotation} and it} may affect the mass ablation during ECI, as well as the effective energy injected in the envelope $E_\mathrm{heat}$, which dictates the magnitude and timescale of the companion's inflation.

Alternative models describing ECI and its subsequent expansion phase, such as those proposed in \cite{Lu25_CCI} and \cite{Chen23_ECI_models}, predict maximum radii compatible with those of \cite{Ogata21_ECI} which are however achieved over years, rather than on a purely dynamical timescale. This would delay the appearance of CCI features in the supernova.

\subsubsection{CCI}\label{sec:disc:ECI_CCI:CCI}
The way we identify whether a post-supernova system undergoes CCI does not guarantee it is observable. The resulting contribution to the light curve (whether a ``bump'' or a periodic modulation) may be too faint or may be masked by other power sources like interaction with the circumstellar medium. This suggests that our estimates of the number of supernovae with CCI are to be taken as upper limits on the observable rate. A rigorous assessment of their observability is left to future work.

\cite{HiraiPosdi22_ECI_NS} and \cite{Hirai25_2022jli_ECI} characterize respectively supernovae undergoing single CCI and periodic CCI, and demonstrate that the NS can drive mass-loss from the companion star as it penetrates the companion’s envelope.
\cite{Hirai25_2022jli_ECI} focused specifically on reproducing \SN{2022jli} and found that the companion star loses less than $0.1\Msun$ within $100\,\days$ through periodic CCI, although these estimates are sensitive to the specific assumptions adopted for radiative feedback as the NS penetrates the envelope.
A more massive compact object (like a BH) or a smaller periastron separation may help enhance the mass loss. This would also efficiently transport away the energy that was injected inside the companion via ECI, which would then shorten the inflation timescale of the companion $\tauinfl$. 

Periodic CCI may reduce the local density of the envelope, diminishing the accretion-powered luminosity and thus the strength of the undulation signal in the light curve \citep{Hirai25_2022jli_ECI}. The observed signal is also also subject to viewing-angle effects and the obscuration from the unbound material \citep{Hirai25_2022jli_ECI}. Consequently, our assumption that periodic CCI is observable for as long as the companion is inflated is optimistic. Finally, post-explosion binaries with higher post-supernova eccentricities or lower periastron distances may yield higher-amplitude undulations in the light curve \citep{Hirai25_2022jli_ECI}, as the NS penetrates deeper into the extended envelope. \EDIT{However, if the NS digs too deep in the envelope, the radiation would diffuse through more of the envelope, smearing out the undulations \citep{Hirai25_2022jli_ECI}.} For sufficiently low eccentricities\EDIT{, on the other hand,} the NS can remain embedded inside the shock-inflated envelope for much of the orbit. The resulting emission would then resemble a quasi-steady power source. In the reference population model (and  those sharing the kicks from \Valli), this would mostly affect the systems that underwent \Case BB RLOF.

\subsection{The post-supernova signatures of CCI}\label{sec:discussion:futureevo}
While we have focused thus far on the immediate effects of CCI on the supernova itself, this interaction may leave behind long-term signatures. These signatures may appear on the NS, the companion star, and the binary orbit, relative to systems in which CCI does not occur.

CCI may affect the companion's spin through the angular-momentum losses \EDIT{via the induced mass-loss} (see Sect.\,\ref{sec:disc:ECI_CCI}). \EDIT{The spin-down} may be significant, since the outer layers that are stripped possess the highest specific angular momentum. \EDIT{If CCI is periodic, the mass and angular momentum losses would be more severe. Concurrently, tidal torques acting during the inflated phase may also become important}. As such, the companion star may either be spun down (potentially yielding an OB star instead of an OBe star), or spun up.  

Furthermore, drag and mass/angular-momentum exchanges during the penetration of the companion's envelope at periastron can reduce both the period and eccentricity. However, this effect may be small, since the undulation period in supernovae such as SN2022jli does not appear to change (\citealt{Zhang25_2022jli_gammarays}, but see \citealt{Maeda25_2022esa_CSI_or_CCI}). The overlap between our models and the observed Be/X-ray binaries compiled by \Valli\ (see the $P_\mathrm{post-SN}-e_\mathrm{post-SN}$ corner plot in Fig.\,\ref{fig:cornerplot_PeN}) suggests that most of the observed systems may have experienced CCI. Consequently, it is possible that the kick distribution of \Valli\ partially incorporates the effect of CCI on the orbit.

\cite{Cary26_CCI_spindown_NS} argue that CCI can \EDIT{slow down the rotation of the NS and produce so-called} ultra-long-period NSs after a single periastron passage. \EDIT{This, however,} depends on the  magnetic field of the NS, which we do not predict. \EDIT{In our models where CCI is periodic, the} repeated interaction may facilitate the spin-down of the NS, increasing the fraction of NSs that evolve into ultra–long-period rotators. 

\subsection{Comparison to previous works}\label{sec:discussion:compare}
\cite{Zapartas25_demographics_companions_of_SESNe} estimate the fraction of non-degenerate companions to H-poor and H-deficient supernovae that undergo significant ECI-induced inflation.
Their population model shares many of the assumptions of our population model \texttt{k3m3e3}, while the underlying stellar evolutionary \EDIT{models} adopt slightly different physics assumption than our own (cf. \citealt{Fragos23_POSYDON, Andrews25_POSYDON2}).

They report that approximately $12\%$ ($5\%$) of the companions to H-poor and H-deficient supernovae are expected to experience ECI-induced radial expansion for more than $5\yr$ ($10\yr$), whereas our comparable population model yields $8\%$ ($2\%$). The difference is mainly methodological. \cite{Zapartas25_demographics_companions_of_SESNe} infer inflation timescales by imposing  $4\pi\tilde\Omega_\mathrm{eff}>0.127$ ($4\pi\tilde\Omega_\mathrm{eff}>0.071$; Eq.\,\ref{eq:OmegaEff}), which is calibrated against a single ECI simulation in \cite{Ogata21_ECI}, while we explicitly compute $\tauinfl$ for each individual companion star. Applying their $\tilde\Omega_\mathrm{eff}$ thresholds to our population results in consistent fractions of $13\%$ ($7\%$).

\section{Conclusions}\label{sec:conclusions}
We analyzed a comprehensive detailed grid of binary- and single-star evolutionary models to quantify the fraction of core-collapse supernovae that experience interaction between the newly born compact-object and a non-degenerate stellar companion, which we referred to as compact-object-companion interaction (CCI). Such interaction has been proposed as the physical origin of periodic modulation of the light curve and H$\alpha$ line observed in events such as \SN{2022jli} \citep{Moore2023_SN2022jli_Ic_Binary, Chen2024_SN2022jli_binary, Hirai25_2022jli_ECI, Zhang25_2022jli_gammarays} and \SN{2015ap} \citep{Ragosta25_2015ap_CCI}.

We performed population synthesis calculations using the binary-model grid, incorporating the effects of the radial expansion of the companion following the impact of the exploding star's ejecta. We find that CCI is predominantly expected in H-poor supernovae (i.e. \Type Ibc and \Type Ibn) while it is rare in H-rich supernovae. This is because supernovae with CCI are predominantly found in tight binary models at the time of the first supernova, which are preferentially H-poor supernovae due to previous phases of mass-transfer. We suggest that similarly \Type IIb supernovae may also exhibit CCI at lower metallicities. Under different assumptions in the population synthesis model (by adopting different kicks, explodability criteria and pre-supernova merger criteria), the predicted fraction of H-poor supernovae exhibiting periodic CCI ranges from $3$ to $27\%$ (Fig.\,\ref{fig:line}). These numbers are to be taken as observational upper limits, as we do not model the observability of the resulting undulations in the light curve. With regard to the post-explosion orbital properties of the supernovae that exhibit periodic CCI, the various population models yield broad orbital period distributions, similarly peaking around $20-50\days$ with increasing eccentricities for models producing larger natal kicks and vice-versa.

We find that the ECI-induced inflation in our models increases the brightness of the companions of H-poor supernovae, which can be 2-200 times higher than their pre-ECI luminosity. This applies generally to all companions, regardless of whether they exhibited CCI or not. These luminosities enhance the probability of observing the companions while they are inflated.   

We find binary evolution models that can match the observed properties of \SN{2022jli} and \SN{2015ap}, all of which have undergone mass-transfer throughout their pre-supernova evolution.
For \SN{2022jli}, the companion may still be inflated and detectable with the VLT's HAWK-I instrument, as we predict a $J$-band magnitude of $21-23$ while \cite{Grosbol12_straformation_GDgalaxies} identified stellar sources close to the location of the supernova position down to $J\sim23$ with a 4-minute exposure. 
We also investigated \SN{2022esa}, which shows a periodic modulation consistent with the orbital periods predicted by our models. However, this transient lacks the characteristic H$\alpha$ emission line found in \SN{2022jli} and \SN{2015ap}, and its variability may be powered by a different physical mechanism (e.g., interaction with concentric shells of circumstellar material, \citealt{Maeda25_2022esa_CSI_or_CCI}).

The models presented here do not provide predictions for the amplitude of the undulating signal in supernovae that undergo periodic CCI, which is essential to assess whether the undulation will be detectable. Nevertheless, our findings suggest that periodic CCI may be much more common than indicated by the few candidates identified. Systematic re-analyzes of archival transients may reveal additional core-collapse supernovae with undulations \citep{Ragosta25_2015ap_CCI, Zhu2026_binariesCCSNeZTF}. High-cadence and deeper time-domain surveys, now enabled by the combined operation of multiple facilities such as the \EDIT{Palomar observatory (running the Zwicky Transient Facility survey) and the Vera C. Rubin observatory (running the Legacy Survey of Space and Time, LSST), as well as future surveys from the Argus Array}, will likely expand the sample of candidates. \EDIT{LSST and future observations from the Nancy Grace Roman Space Telescope will particularly be useful for detecting inflated supernova companions in late-time imaging}. As the sample of identified transients increases, it will become possible to place constraints on the physical processes governing binary evolution and supernova explosions.

\begin{acknowledgements}
This project made use of the Julia language \citep{code_Julia2017}. The authors are grateful to the anonymous referee for helpful and constructive comments that improved the quality of the manuscript. A.E. also thanks Ryosuke Hirai, Philipp Podsiadlowski, Steve Schulze, Manos Zapartas, Dimitrios Souropanis, Sylvia Zhu, Thomas Moore, Ori Fox, Thomas Tauris, Hans-Thomas Janka, Raphael Baer-Way, and Jan R\"order for interesting discussions.  A.E. acknowledges the support from the DFG through grant LA 587/22-1.

\end{acknowledgements}

\bibliographystyle{aa}
\bibliography{Reference_list}

\begin{thebibliography}{144}
\expandafter\ifx\csname natexlab\endcsname\relax\def\natexlab#1{#1}\fi

\bibitem[{{Almeida} {et~al.}(2017){Almeida}, {Sana}, {Taylor}, {Barb{\'a}}, {Bonanos}, {Crowther}, {Damineli}, {de Koter}, {de Mink}, {Evans}, {Gieles}, {Grin}, {H{\'e}nault-Brunet}, {Langer}, {Lennon}, {Lockwood}, {Ma{\'\i}z Apell{\'a}niz}, {Moffat}, {Neijssel}, {Norman}, {Ram{\'\i}rez-Agudelo}, {Richardson}, {Schootemeijer}, {Shenar}, {Soszy{\'n}ski}, {Tramper}, \& {Vink}}]{Almeida17_FLAMES_OB_binaries}
{Almeida}, L.~A., {Sana}, H., {Taylor}, W., {et~al.} 2017, \aap, 598, A84

\bibitem[{{Andrews} {et~al.}(2025){Andrews}, {Bavera}, {Briel}, {Chattaraj}, {Dotter}, {Fragos}, {Gallegos-Garcia}, {Gossage}, {Kalogera}, {Kasdagli}, {Katsaggelos}, {Kimball}, {Kovlakas}, {Kruckow}, {Liotine}, {Misra}, {Rocha}, {Souropanis}, {Srivastava}, {Sun}, {Teng}, {Xing}, {Zapartas}, \& {Zevin}}]{Andrews25_POSYDON2}
{Andrews}, J.~J., {Bavera}, S.~S., {Briel}, M., {et~al.} 2025, \apjs, 281, 3

\bibitem[{{Angus} {et~al.}(2024){Angus}, {Woosley}, {Foley}, {Nicholl}, {Villar}, {Taggart}, {Pursiainen}, {Ramsden}, {Srivastav}, {Stevance}, {Moore}, {Auchettl}, {Hoogendam}, {Khetan}, {Yadavalli}, {Dimitriadis}, {Gagliano}, {Siebert}, {Aamer}, {de Boer}, {Chambers}, {Clocchiatti}, {Coulter}, {Drout}, {Farias}, {Fulton}, {Gall}, {Gao}, {Izzo}, {Jones}, {Lin}, {Magnier}, {Narayan}, {Ramirez-Ruiz}, {Ransome}, {Rest}, {Smartt}, \& {Smith}}]{Angus24_2024acct_doublepeaked}
{Angus}, C.~R., {Woosley}, S.~E., {Foley}, R.~J., {et~al.} 2024, \apjl, 977, L41

\bibitem[{{Arcavi} {et~al.}(2017){Arcavi}, {Howell}, {Kasen}, {Bildsten}, {Hosseinzadeh}, {McCully}, {Wong}, {Katz}, {Gal-Yam}, {Sollerman}, {Taddia}, {Leloudas}, {Fremling}, {Nugent}, {Horesh}, {Mooley}, {Rumsey}, {Cenko}, {Graham}, {Perley}, {Nakar}, {Shaviv}, {Bromberg}, {Shen}, {Ofek}, {Cao}, {Wang}, {Huang}, {Rui}, {Zhang}, {Li}, {Li}, {Zhang}, {Valenti}, {Guevel}, {Shappee}, {Kochanek}, {Holoien}, {Filippenko}, {Fender}, {Nyholm}, {Yaron}, {Kasliwal}, {Sullivan}, {Blagorodnova}, {Walters}, {Lunnan}, {Khazov}, {Andreoni}, {Laher}, {Konidaris}, {Wozniak}, \& {Bue}}]{Arcavi17_iPTF14hls}
{Arcavi}, I., {Howell}, D.~A., {Kasen}, D., {et~al.} 2017, \nat, 551, 210

\bibitem[{{Aryan} {et~al.}(2021){Aryan}, {Pandey}, {Zheng}, {Filippenko}, {Vinko}, {Ouchi}, {Shivvers}, {Yuk}, {Kumar}, {Stegman}, {Halevi}, {Ross}, {Gould}, {Yunus}, {Baer-Way}, {deGraw}, {Maeda}, {Bhattacharya}, {Kumar}, {Gupta}, {Yadav}, {Buckley}, {Misra}, \& {Tiwari}}]{Aryan21_sn2015ap_sn2016bau_progenitor}
{Aryan}, A., {Pandey}, S.~B., {Zheng}, W., {et~al.} 2021, \mnras, 505, 2530

\bibitem[{{Baer-way} {et~al.}(2026){Baer-way}, {Chandra}, {Modjaz}, {Nayana}, {Maeda}, {Auchettl}, {Drout}, {Kilpatrick}, {Ray}, \& {Ryder}}]{BaerWay2026_Sn2019yvr_lateH_CSM}
{Baer-way}, R., {Chandra}, P., {Modjaz}, M., {et~al.} 2026, arXiv e-prints, arXiv:2607.05500

\bibitem[{Bezanson {et~al.}(2017)Bezanson, Edelman, Karpinski, \& Shah}]{code_Julia2017}
Bezanson, J., Edelman, A., Karpinski, S., \& Shah, V.~B. 2017, SIAM review, 59, 65

\bibitem[{{Blaauw}(1961)}]{BlaauwKick_1961}
{Blaauw}, A. 1961, \bain, 15, 265

\bibitem[{{Blinnikov} \& {Bartunov}(1993)}]{Blinnikov1993_STELLA}
{Blinnikov}, S.~I. \& {Bartunov}, O.~S. 1993, \aap, 273, 106

\bibitem[{{Bondi} \& {Hoyle}(1944)}]{BondiHoyle44_accretion}
{Bondi}, H. \& {Hoyle}, F. 1944, \mnras, 104, 273

\bibitem[{{Brennan} {et~al.}(2025){Brennan}, {Barmentloo}, {Schulze}, {Smith}, {Hirai}, {Eldridge}, {Fraser}, {Stevance}, {Smartt}, {Anand}, {Aryan}, {Chen}, {Das}, {Drake}, {Fransson}, {Gangopadhyay}, {Gkini}, {Jacobson-Gal\textbackslash'an}, {Jerkstrand}, {Johansson}, {Nicholl}, {Pignata}, {Sarin}, {Singh}, {Sollerman}, {Srivastav}, {van Baal}, {Chambers}, {Coughlin}, {Gao}, {Graham}, {Huber}, {Lin}, {Lowe}, {Magnier}, {Masci}, {Purdum}, {Rest}, {Rusholme}, {Smith}, {Smith}, {Tweddle}, {Wainscoat}, \& {de Boer}}]{Brennan2025_SN2022mop}
{Brennan}, S.~J., {Barmentloo}, S., {Schulze}, S., {et~al.} 2025, arXiv e-prints, arXiv:2503.08768

\bibitem[{{Brott} {et~al.}(2011){Brott}, {de Mink}, {Cantiello}, {Langer}, {de Koter}, {Evans}, {Hunter}, {Trundle}, \& {Vink}}]{Brott2011}
{Brott}, I., {de Mink}, S.~E., {Cantiello}, M., {et~al.} 2011, \aap, 530, A115

\bibitem[{{Burrows} {et~al.}(2025){Burrows}, {Wang}, \& {Vartanyan}}]{Burrows25_BH_formation_CCSN}
{Burrows}, A., {Wang}, T., \& {Vartanyan}, D. 2025, \apj, 987, 164

\bibitem[{{Burrows} {et~al.}(2024){Burrows}, {Wang}, {Vartanyan}, \& {Coleman}}]{Burrows24_NS_BH_kicks}
{Burrows}, A., {Wang}, T., {Vartanyan}, D., \& {Coleman}, M. S.~B. 2024, \apj, 963, 63

\bibitem[{{Cardelli} {et~al.}(1989){Cardelli}, {Clayton}, \& {Mathis}}]{Cardelli1989_extinction}
{Cardelli}, J.~A., {Clayton}, G.~C., \& {Mathis}, J.~S. 1989, \apj, 345, 245

\bibitem[{{Cartier} {et~al.}(2026){Cartier}, {Contreras}, {Stritzinger}, {Hamuy}, {Ruiz-Lapuente}, {Prieto}, {Anderson}, {Cikota}, \& {Gerlach}}]{Cartier26_2022jli_magnetar}
{Cartier}, R., {Contreras}, C., {Stritzinger}, M., {et~al.} 2026, \aap, 707, A161

\bibitem[{{Cary} {et~al.}(2026){Cary}, {Lu}, {Leung}, \& {Wong}}]{Cary26_CCI_spindown_NS}
{Cary}, S., {Lu}, W., {Leung}, C., \& {Wong}, T. L.~S. 2026, \apj, 996, 141

\bibitem[{{Chan} {et~al.}(2020){Chan}, {M{\"u}ller}, \& {Heger}}]{Chan2020_BH_formation_from_CCSNe_fallback}
{Chan}, C., {M{\"u}ller}, B., \& {Heger}, A. 2020, \mnras, 495, 3751

\bibitem[{{Chan} {et~al.}(2018){Chan}, {M{\"u}ller}, {Heger}, {Pakmor}, \& {Springel}}]{Chan2018_ccsn_explosion_fallback_BH}
{Chan}, C., {M{\"u}ller}, B., {Heger}, A., {Pakmor}, R., \& {Springel}, V. 2018, \apjl, 852, L19

\bibitem[{{Chatzopoulos} {et~al.}(2012){Chatzopoulos}, {Wheeler}, \& {Vinko}}]{Chatzopoulos12_CSM_SLSN}
{Chatzopoulos}, E., {Wheeler}, J.~C., \& {Vinko}, J. 2012, \apj, 746, 121

\bibitem[{{Chen} {et~al.}(2023{\natexlab{a}}){Chen}, {Rau}, \& {Pan}}]{Chen23_ECI_models}
{Chen}, H.-P., {Rau}, S.-J., \& {Pan}, K.-C. 2023{\natexlab{a}}, \apj, 949, 121

\bibitem[{{Chen} {et~al.}(2024){Chen}, {Gal-Yam}, {Sollerman}, {Schulze}, {Post}, {Liu}, {Ofek}, {Das}, {Fremling}, {Horesh}, {Katz}, {Kushnir}, {Kasliwal}, {Kulkarni}, {Liu}, {Liu}, {Miller}, {Rose}, {Waxman}, {Yang}, {Yao}, {Zackay}, {Bellm}, {Dekany}, {Drake}, {Fang}, {Fynbo}, {Groom}, {Helou}, {Irani}, {Jegou du Laz}, {Liu}, {Mazzali}, {Neill}, {Qin}, {Riddle}, {Sharon}, {Strotjohann}, {Wold}, \& {Yan}}]{Chen2024_SN2022jli_binary}
{Chen}, P., {Gal-Yam}, A., {Sollerman}, J., {et~al.} 2024, \nat, 625, 253

\bibitem[{{Chen} {et~al.}(2023{\natexlab{b}}){Chen}, {Yan}, {Kangas}, {Lunnan}, {Sollerman}, {Schulze}, {Perley}, {Chen}, {Taggart}, {Hinds}, {Gal-Yam}, {Wang}, {De}, {Bellm}, {Bloom}, {Dekany}, {Graham}, {Kasliwal}, {Kulkarni}, {Laher}, {Neill}, \& {Rusholme}}]{Chen2023_SLSNEI_CSM_Magnetar_undulations}
{Chen}, Z.~H., {Yan}, L., {Kangas}, T., {et~al.} 2023{\natexlab{b}}, \apj, 943, 42

\bibitem[{{Chevalier}(2012)}]{Chevalier2012_mergerburst}
{Chevalier}, R.~A. 2012, \apjl, 752, L2

\bibitem[{{Choi} {et~al.}(2016){Choi}, {Dotter}, {Conroy}, {Cantiello}, {Paxton}, \& {Johnson}}]{Choi16_MIST}
{Choi}, J., {Dotter}, A., {Conroy}, C., {et~al.} 2016, \apj, 823, 102

\bibitem[{{de Kool}(1990)}]{deKool_CEE}
{de Kool}, M. 1990, \apj, 358, 189

\bibitem[{{Dessart}(2018)}]{Dessart2018_iPTF14hls}
{Dessart}, L. 2018, \aap, 610, L10

\bibitem[{{Dessart} {et~al.}(2012){Dessart}, {Hillier}, {Li}, \& {Woosley}}]{Dessart2012_IbcSNe}
{Dessart}, L., {Hillier}, D.~J., {Li}, C., \& {Woosley}, S. 2012, \mnras, 424, 2139

\bibitem[{{Dessart} {et~al.}(2016){Dessart}, {Hillier}, {Woosley}, {Livne}, {Waldman}, {Yoon}, \& {Langer}}]{Dessart2016_IIb_Ibc_RadTransf2}
{Dessart}, L., {Hillier}, D.~J., {Woosley}, S., {et~al.} 2016, \mnras, 458, 1618

\bibitem[{{Dessart} {et~al.}(2020{\natexlab{a}}){Dessart}, {Leonard}, \& {Prieto}}]{Dessart2020_spectraIa}
{Dessart}, L., {Leonard}, D.~C., \& {Prieto}, J.~L. 2020{\natexlab{a}}, \aap, 638, A80

\bibitem[{{Dessart} {et~al.}(2020{\natexlab{b}}){Dessart}, {Yoon}, {Aguilera-Dena}, \& {Langer}}]{Dessart2020_Ibc}
{Dessart}, L., {Yoon}, S.-C., {Aguilera-Dena}, D.~R., \& {Langer}, N. 2020{\natexlab{b}}, \aap, 642, A106

\bibitem[{{Dewi} \& {Tauris}(2000)}]{DewiTauris_CEE}
{Dewi}, J.~D.~M. \& {Tauris}, T.~M. 2000, \aap, 360, 1043

\bibitem[{{Dexter} \& {Kasen}(2013)}]{DexterKasen13_FallbackAccretion_SNe}
{Dexter}, J. \& {Kasen}, D. 2013, \apj, 772, 30

\bibitem[{{Disberg} \& {Mandel}(2025)}]{DM25_kicks_isolated_ns}
{Disberg}, P. \& {Mandel}, I. 2025, \apjl, 989, L8

\bibitem[{{Dotter}(2016)}]{Dotter16_MIST}
{Dotter}, A. 2016, \apjs, 222, 8

\bibitem[{{Eldridge} {et~al.}(2013){Eldridge}, {Fraser}, {Smartt}, {Maund}, \& {Crockett}}]{Eldridge13_deathmassivestarII_Ibc}
{Eldridge}, J.~J., {Fraser}, M., {Smartt}, S.~J., {Maund}, J.~R., \& {Crockett}, R.~M. 2013, \mnras, 436, 774

\bibitem[{{Ercolino} {et~al.}(2024){Ercolino}, {Jin}, {Langer}, \& {Dessart}}]{Ercolino_widebinary_RSG}
{Ercolino}, A., {Jin}, H., {Langer}, N., \& {Dessart}, L. 2024, \aap, 685, A58

\bibitem[{{Ercolino} {et~al.}(2025){Ercolino}, {Jin}, {Langer}, \& {Dessart}}]{Ercolino_HpoorInteractingSNe}
{Ercolino}, A., {Jin}, H., {Langer}, N., \& {Dessart}, L. 2025, \aap, 696, A103

\bibitem[{{Ercolino} {et~al.}(2026){Ercolino}, {Jin}, {Langer}, {Gal-Yam}, {Schootemeijer}, \& {Mannes}}]{Ercolino25_popsynth_CCSNE_I}
{Ercolino}, A., {Jin}, H., {Langer}, N., {et~al.} 2026, \aap, 706, A169

\bibitem[{{Ertl} {et~al.}(2016){Ertl}, {Janka}, {Woosley}, {Sukhbold}, \& {Ugliano}}]{Ertl16_explodability_m4mu4}
{Ertl}, T., {Janka}, H.~T., {Woosley}, S.~E., {Sukhbold}, T., \& {Ugliano}, M. 2016, \apj, 818, 124

\bibitem[{{Ertl} {et~al.}(2020){Ertl}, {Woosley}, {Sukhbold}, \& {Janka}}]{Ertl2020_explodability_HeS}
{Ertl}, T., {Woosley}, S.~E., {Sukhbold}, T., \& {Janka}, H.~T. 2020, \apj, 890, 51

\bibitem[{{Farah} {et~al.}(2026){Farah}, {Prust}, {Howell}, {Ni}, {McCully}, {Andrews}, {Kumar}, {Hiramatsu}, {Gomez}, {Wynn}, {Filippenko}, {Bostroem}, {Berger}, \& {Blanchard}}]{Farah25_SLSNI_undulations_magnetar}
{Farah}, J.~R., {Prust}, L.~J., {Howell}, D.~A., {et~al.} 2026, \nat, 651, 321

\bibitem[{{Ferrari} {et~al.}(2024){Ferrari}, {Folatelli}, {Kuncarayakti}, {Stritzinger}, {Maeda}, {Bersten}, {Rom{\'a}n Aguilar}, {S{\'a}ez}, {Dessart}, {Lundqvist}, {Mazzali}, {Nagao}, {Ashall}, {Bose}, {Brennan}, {Cai}, {Handberg}, {Holmbo}, {Karamehmetoglu}, {Pastorello}, {Reguitti}, {Anderson}, {Chen}, {Galbany}, {Gromadzki}, {Guti{\'e}rrez}, {Inserra}, {Kankare}, {M{\"u}ller Bravo}, {Mattila}, {Nicholl}, {Pignata}, {Sollerman}, {Srivastav}, \& {Young}}]{Ferrari2024_Sn2019yvr_latetimeH_interaction}
{Ferrari}, L., {Folatelli}, G., {Kuncarayakti}, H., {et~al.} 2024, \mnras, 529, L33

\bibitem[{{Fox} {et~al.}(2014){Fox}, {Azalee Bostroem}, {Van Dyk}, {Filippenko}, {Fransson}, {Matheson}, {Cenko}, {Chandra}, {Dwarkadas}, {Li}, {Parker}, \& {Smith}}]{Fox14_companion1993J}
{Fox}, O.~D., {Azalee Bostroem}, K., {Van Dyk}, S.~D., {et~al.} 2014, \apj, 790, 17

\bibitem[{{Fox} {et~al.}(2022){Fox}, {Van Dyk}, {Williams}, {Drout}, {Zapartas}, {Smith}, {Milisavljevic}, {Andrews}, {Bostroem}, {Filippenko}, {Gomez}, {Kelly}, {de Mink}, {Pierel}, {Rest}, {Ryder}, {Sravan}, {Strolger}, {Wang}, \& {Weil}}]{Fox2022_companion_2013ge}
{Fox}, O.~D., {Van Dyk}, S.~D., {Williams}, B.~F., {et~al.} 2022, \apjl, 929, L15

\bibitem[{{Fragos} {et~al.}(2023){Fragos}, {Andrews}, {Bavera}, {Berry}, {Coughlin}, {Dotter}, {Giri}, {Kalogera}, {Katsaggelos}, {Kovlakas}, \& et~al.}]{Fragos23_POSYDON}
{Fragos}, T., {Andrews}, J.~J., {Bavera}, S.~S., {et~al.} 2023, \apjs, 264, 45

\bibitem[{{Gal-Yam}(2017)}]{GalYam17_SN_Book}
{Gal-Yam}, A. 2017, in Handbook of Supernovae, ed. A.~W. {Alsabti} \& P.~{Murdin}, 195

\bibitem[{{Grosb{\o}l} \& {Dottori}(2012)}]{Grosbol12_straformation_GDgalaxies}
{Grosb{\o}l}, P. \& {Dottori}, H. 2012, \aap, 542, A39

\bibitem[{{Harrison} \& {Tademaru}(1975)}]{HarrisonTademaru75_NSRocket}
{Harrison}, E.~R. \& {Tademaru}, E. 1975, \apj, 201, 447

\bibitem[{{Hirai}(2023)}]{Hirai23_ECI}
{Hirai}, R. 2023, \mnras, 523, 6011

\bibitem[{{Hirai} \& {Podsiadlowski}(2022)}]{HiraiPosdi22_ECI_NS}
{Hirai}, R. \& {Podsiadlowski}, P. 2022, \mnras, 517, 4544

\bibitem[{{Hirai} {et~al.}(2024){Hirai}, {Podsiadlowski}, {Heger}, \& {Nagakura}}]{Hirai24_NSRocket}
{Hirai}, R., {Podsiadlowski}, P., {Heger}, A., \& {Nagakura}, H. 2024, \apjl, 972, L18

\bibitem[{{Hirai} {et~al.}(2025){Hirai}, {Podsiadlowski}, {Hoeflich}, {Barkov}, {Chan}, {Liptai}, \& {Nagataki}}]{Hirai25_2022jli_ECI}
{Hirai}, R., {Podsiadlowski}, P., {Hoeflich}, P., {et~al.} 2025, \apj, 995, 55

\bibitem[{{Hirai} {et~al.}(2018){Hirai}, {Podsiadlowski}, \& {Yamada}}]{Hirai18_ECI}
{Hirai}, R., {Podsiadlowski}, P., \& {Yamada}, S. 2018, \apj, 864, 119

\bibitem[{{Hirai} {et~al.}(2014){Hirai}, {Sawai}, \& {Yamada}}]{Hirai14_CCSNe_ejecta_impact_companion}
{Hirai}, R., {Sawai}, H., \& {Yamada}, S. 2014, \apj, 792, 66

\bibitem[{{Hirai} \& {Yamada}(2015)}]{HiraiYamada15_ECI}
{Hirai}, R. \& {Yamada}, S. 2015, \apj, 805, 170

\bibitem[{{Hobbs} {et~al.}(2005){Hobbs}, {Lorimer}, {Lyne}, \& {Kramer}}]{Hobbs2005_kicks_isolated_pulsars}
{Hobbs}, G., {Lorimer}, D.~R., {Lyne}, A.~G., \& {Kramer}, M. 2005, \mnras, 360, 974

\bibitem[{{Hosseinzadeh} {et~al.}(2022){Hosseinzadeh}, {Berger}, {Metzger}, {Gomez}, {Nicholl}, \& {Blanchard}}]{Hosseinzadeh22_SLSNe_bumps_and_oscillations}
{Hosseinzadeh}, G., {Berger}, E., {Metzger}, B.~D., {et~al.} 2022, \apj, 933, 14

\bibitem[{{Humphreys} \& {Davidson}(1979)}]{HumphreyDavidson1979}
{Humphreys}, R.~M. \& {Davidson}, K. 1979, \apj, 232, 409

\bibitem[{{Janka}(2013)}]{Janka2013_kicks_BHs}
{Janka}, H.-T. 2013, \mnras, 434, 1355

\bibitem[{{Janka} \& {Kresse}(2024)}]{JankaKresse24_NS_BH_kicks}
{Janka}, H.-T. \& {Kresse}, D. 2024, \apss, 369, 80

\bibitem[{{Jin} {et~al.}(2026){Jin}, {Langer}, {Ercolino}, \& {de Mink}}]{Jin25_BonnGal}
{Jin}, H., {Langer}, N., {Ercolino}, A., \& {de Mink}, S.~E. 2026, \aap, 707, A56

\bibitem[{{Jin} {et~al.}(2024){Jin}, {Langer}, {Lennon}, \& {Proffitt}}]{Jin2024_boron}
{Jin}, H., {Langer}, N., {Lennon}, D.~J., \& {Proffitt}, C.~R. 2024, \aap, 690, A135

\bibitem[{{Kasen}(2010)}]{Kasen10_ejecta_ccsn_impact_sec}
{Kasen}, D. 2010, \apj, 708, 1025

\bibitem[{{Kasen} \& {Bildsten}(2010)}]{KasenBildstein2010_MagnetarSpindown_poweringSNe}
{Kasen}, D. \& {Bildsten}, L. 2010, \apj, 717, 245

\bibitem[{{Kilpatrick} {et~al.}(2021){Kilpatrick}, {Drout}, {Auchettl}, {Dimitriadis}, {Foley}, {Jones}, {DeMarchi}, {French}, {Gall}, {Hjorth}, {Jacobson-Gal{\'a}n}, {Margutti}, {Piro}, {Ramirez-Ruiz}, {Rest}, \& {Rojas-Bravo}}]{Kilpatrick2021_2019yvr_latetimeinteraction}
{Kilpatrick}, C.~D., {Drout}, M.~R., {Auchettl}, K., {et~al.} 2021, \mnras, 504, 2073

\bibitem[{{Kruckow} {et~al.}(2018){Kruckow}, {Tauris}, {Langer}, {Kramer}, \& {Izzard}}]{Kruckow18_COMBINE}
{Kruckow}, M.~U., {Tauris}, T.~M., {Langer}, N., {Kramer}, M., \& {Izzard}, R.~G. 2018, \mnras, 481, 1908

\bibitem[{{Kumar} {et~al.}(2026){Kumar}, {Blanchard}, {Berger}, {Athukoralalage}, {Hiramatsu}, {Gomez}, {Andrews}, {Bostroem}, {Farah}, {Howell}, \& {McCully}}]{Kumar25_2024afav_SLSNI_with_undulations}
{Kumar}, H., {Blanchard}, P.~K., {Berger}, E., {et~al.} 2026, \apjl, 998, L3

\bibitem[{{Kuncarayakti} {et~al.}(2023){Kuncarayakti}, {Sollerman}, {Izzo}, {Maeda}, {Yang}, {Schulze}, {Angus}, {Aubert}, {Auchettl}, {Della Valle}, {Dessart}, {Hinds}, {Kankare}, {Kawabata}, {Lundqvist}, {Nakaoka}, {Perley}, {Raimundo}, {Strotjohann}, {Taguchi}, {Cai}, {Charalampopoulos}, {Fang}, {Fraser}, {Guti{\'e}rrez}, {Imazawa}, {Kangas}, {Kawabata}, {Kotak}, {Kravtsov}, {Matilainen}, {Mattila}, {Moran}, {Murata}, {Salmaso}, {Anderson}, {Ashall}, {Bellm}, {Benetti}, {Chambers}, {Chen}, {Coughlin}, {De Colle}, {Fremling}, {Galbany}, {Gal-Yam}, {Gromadzki}, {Groom}, {Hajela}, {Inserra}, {Kasliwal}, {Mahabal}, {Martin-Carrillo}, {Moore}, {M{\"u}ller-Bravo}, {Nicholl}, {Ragosta}, {Riddle}, {Sharma}, {Srivastav}, {Stritzinger}, {Wold}, \& {Young}}]{Kuncarayakti2023_SN2022xxf_IcBL_multipeak}
{Kuncarayakti}, H., {Sollerman}, J., {Izzo}, L., {et~al.} 2023, \aap, 678, A209

\bibitem[{{Langer}(2012)}]{Langer_review_2012}
{Langer}, N. 2012, \araa, 50, 107

\bibitem[{{Lechien} {et~al.}(2025){Lechien}, {de Mink}, {Valli}, {Rubio}, {van Son}, {Klement}, {Jin}, \& {Pols}}]{Lechien25_accretionefficiency}
{Lechien}, T., {de Mink}, S.~E., {Valli}, R., {et~al.} 2025, \apjl, 990, L51

\bibitem[{{Liu} {et~al.}(2015{\natexlab{a}}){Liu}, {Moriya}, \& {Stancliffe}}]{Liu15_UV_Ia_signatures_companion}
{Liu}, Z.-W., {Moriya}, T.~J., \& {Stancliffe}, R.~J. 2015{\natexlab{a}}, \mnras, 454, 1192

\bibitem[{{Liu} {et~al.}(2015{\natexlab{b}}){Liu}, {Tauris}, {R{\"o}pke}, {Moriya}, {Kruckow}, {Stancliffe}, \& {Izzard}}]{Liu15_ccsne_ejecta_interaction_with_ms}
{Liu}, Z.-W., {Tauris}, T.~M., {R{\"o}pke}, F.~K., {et~al.} 2015{\natexlab{b}}, \aap, 584, A11

\bibitem[{{Lu} {et~al.}(2025){Lu}, {Cary}, \& {Tsuna}}]{Lu25_CCI}
{Lu}, W., {Cary}, S., \& {Tsuna}, D. 2025, arXiv e-prints, arXiv:2507.14284

\bibitem[{{Maeda} {et~al.}(2026){Maeda}, {Kuncarayakti}, {Nagao}, {Kawabata}, {Taguchi}, {Uno}, \& {De}}]{Maeda25_2022esa_CSI_or_CCI}
{Maeda}, K., {Kuncarayakti}, H., {Nagao}, T., {et~al.} 2026, \pasj, 78, L1

\bibitem[{{Maeda} {et~al.}(2007){Maeda}, {Tanaka}, {Nomoto}, {Tominaga}, {Kawabata}, {Mazzali}, {Umeda}, {Suzuki}, \& {Hattori}}]{Maeda2007_SN2005bf_Ib_magnetarspindown}
{Maeda}, K., {Tanaka}, M., {Nomoto}, K., {et~al.} 2007, \apj, 666, 1069

\bibitem[{{Marchant}(2017)}]{PABLO_LMC_GRID}
{Marchant}, P. 2017, PhD thesis, Rheinische Friedrich Wilhelms University of Bonn, Germany

\bibitem[{{Marietta} {et~al.}(2000){Marietta}, {Burrows}, \& {Fryxell}}]{Marietta2000_Ia_comapnion_ms_2DHD}
{Marietta}, E., {Burrows}, A., \& {Fryxell}, B. 2000, \apjs, 128, 615

\bibitem[{{Maund}(2019)}]{Maund2019_companion2011dh}
{Maund}, J.~R. 2019, \apj, 883, 86

\bibitem[{{Maund} {et~al.}(2004){Maund}, {Smartt}, {Kudritzki}, {Podsiadlowski}, \& {Gilmore}}]{Maund2004_1993J}
{Maund}, J.~R., {Smartt}, S.~J., {Kudritzki}, R.~P., {Podsiadlowski}, P., \& {Gilmore}, G.~F. 2004, \nat, 427, 129

\bibitem[{{Moe} \& {Di Stefano}(2017)}]{MoeDiStefano2017}
{Moe}, M. \& {Di Stefano}, R. 2017, \apjs, 230, 15

\bibitem[{{Moore} {et~al.}(2023){Moore}, {Smartt}, {Nicholl}, {Srivastav}, {Stevance}, {Jess}, {Grant}, {Fulton}, {Rhodes}, {Sim}, {Hirai}, {Podsiadlowski}, {Anderson}, {Ashall}, {Bate}, {Fender}, {Guti{\'e}rrez}, {Howell}, {Huber}, {Inserra}, {Leloudas}, {Monard}, {M{\"u}ller-Bravo}, {Shappee}, {Smith}, {Terreran}, {Tonry}, {Tucker}, {Young}, {Aamer}, {Chen}, {Ragosta}, {Galbany}, {Gromadzki}, {Harvey}, {Hoeflich}, {McCully}, {Newsome}, {Gonzalez}, {Pellegrino}, {Ramsden}, {P{\'e}rez-Torres}, {Ridley}, {Sheng}, \& {Weston}}]{Moore2023_SN2022jli_Ic_Binary}
{Moore}, T., {Smartt}, S.~J., {Nicholl}, M., {et~al.} 2023, \apjl, 956, L31

\bibitem[{{Moriya} {et~al.}(2011){Moriya}, {Tominaga}, {Blinnikov}, {Baklanov}, \& {Sorokina}}]{Moriya11_CSM_SNe_RSG}
{Moriya}, T., {Tominaga}, N., {Blinnikov}, S.~I., {Baklanov}, P.~V., \& {Sorokina}, E.~I. 2011, \mnras, 415, 199

\bibitem[{{Morozova} {et~al.}(2015){Morozova}, {Piro}, {Renzo}, {Ott}, {Clausen}, {Couch}, {Ellis}, \& {Roberts}}]{Morozova15_LCofCCSNe_SNEC}
{Morozova}, V., {Piro}, A.~L., {Renzo}, M., {et~al.} 2015, \apj, 814, 63

\bibitem[{{Mould} {et~al.}(2000){Mould}, {Huchra}, {Freedman}, {Kennicutt}, {Ferrarese}, {Ford}, {Gibson}, {Graham}, {Hughes}, {Illingworth}, {Kelson}, {Macri}, {Madore}, {Sakai}, {Sebo}, {Silbermann}, \& {Stetson}}]{Mould2000_HST_extragalactic_distances}
{Mould}, J.~R., {Huchra}, J.~P., {Freedman}, W.~L., {et~al.} 2000, \apj, 529, 786

\bibitem[{{M{\"u}ller} {et~al.}(2016){M{\"u}ller}, {Heger}, {Liptai}, \& {Cameron}}]{MHLC16}
{M{\"u}ller}, B., {Heger}, A., {Liptai}, D., \& {Cameron}, J.~B. 2016, \mnras, 460, 742

\bibitem[{{Nagarajan} \& {El-Badry}(2025)}]{Nagarajan_ElBadry24_natalkicks_for_BHs}
{Nagarajan}, P. \& {El-Badry}, K. 2025, Publications of the Astronomical Society of the Pacific, 137, 034203

\bibitem[{{Nicholl} {et~al.}(2016){Nicholl}, {Berger}, {Smartt}, {Margutti}, {Kamble}, {Alexander}, {Chen}, {Inserra}, {Arcavi}, {Blanchard}, {Cartier}, {Chambers}, {Childress}, {Chornock}, {Cowperthwaite}, {Drout}, {Flewelling}, {Fraser}, {Gal-Yam}, {Galbany}, {Harmanen}, {Holoien}, {Hosseinzadeh}, {Howell}, {Huber}, {Jerkstrand}, {Kankare}, {Kochanek}, {Lin}, {Lunnan}, {Magnier}, {Maguire}, {McCully}, {McDonald}, {Metzger}, {Milisavljevic}, {Mitra}, {Reynolds}, {Saario}, {Shappee}, {Smith}, {Valenti}, {Villar}, {Waters}, \& {Young}}]{Nicholl2016_2015bn_SLSN_undulation}
{Nicholl}, M., {Berger}, E., {Smartt}, S.~J., {et~al.} 2016, \apj, 826, 39

\bibitem[{{Nyholm} {et~al.}(2017){Nyholm}, {Sollerman}, {Taddia}, {Fremling}, {Moriya}, {Ofek}, {Gal-Yam}, {De Cia}, {Roy}, {Kasliwal}, {Cao}, {Nugent}, \& {Masci}}]{Nyholm17_iPTF13z_IIn_with_undulations_bump}
{Nyholm}, A., {Sollerman}, J., {Taddia}, F., {et~al.} 2017, \aap, 605, A6

\bibitem[{{Nyholm} {et~al.}(2020){Nyholm}, {Sollerman}, {Tartaglia}, {Taddia}, {Fremling}, {Blagorodnova}, {Filippenko}, {Gal-Yam}, {Howell}, {Karamehmetoglu}, {Kulkarni}, {Laher}, {Leloudas}, {Masci}, {Kasliwal}, {Mor{\r{a}}}, {Moriya}, {Ofek}, {Papadogiannakis}, {Quimby}, {Rebbapragada}, \& {Schulze}}]{Nyholm2020_IInSNe_bumps}
{Nyholm}, A., {Sollerman}, J., {Tartaglia}, L., {et~al.} 2020, \aap, 637, A73

\bibitem[{{Ogata} {et~al.}(2021){Ogata}, {Hirai}, \& {Hijikawa}}]{Ogata21_ECI}
{Ogata}, M., {Hirai}, R., \& {Hijikawa}, K. 2021, \mnras, 505, 2485

\bibitem[{{Orellana} {et~al.}(2025){Orellana}, {Bersten}, \& {Guti{\'e}rrez}}]{Orellana25_2022jli_magnetar}
{Orellana}, M., {Bersten}, M.~C., \& {Guti{\'e}rrez}, C.~P. 2025, \aap, 700, L17

\bibitem[{{Pakmor} {et~al.}(2008){Pakmor}, {R{\"o}pke}, {Weiss}, \& {Hillebrandt}}]{Pakmor08_Ia_ejecta_vs_ms}
{Pakmor}, R., {R{\"o}pke}, F.~K., {Weiss}, A., \& {Hillebrandt}, W. 2008, \aap, 489, 943

\bibitem[{{Pan} {et~al.}(2012){Pan}, {Ricker}, \& {Taam}}]{Pan2012_ECI_slowdown_companionrotation}
{Pan}, K.-C., {Ricker}, P.~M., \& {Taam}, R.~E. 2012, \apj, 750, 151

\bibitem[{{Patton} \& {Sukhbold}(2020)}]{PS20_explodability_Mco_Xc}
{Patton}, R.~A. \& {Sukhbold}, T. 2020, \mnras, 499, 2803

\bibitem[{{Pauli}(2020)}]{Pauli_MasterThesis}
{Pauli}, D. 2020, Master's thesis, Rheinische Friedrich Wilhelms University of Bonn, Germany

\bibitem[{{Pavlovskii} \& {Ivanova}(2015)}]{Pavlovskii_Ivanova_2015_MT_from_Giants}
{Pavlovskii}, K. \& {Ivanova}, N. 2015, \mnras, 449, 4415

\bibitem[{{Paxton} {et~al.}(2011){Paxton}, {Bildsten}, {Dotter}, {Herwig}, {Lesaffre}, \& {Timmes}}]{MESA_I}
{Paxton}, B., {Bildsten}, L., {Dotter}, A., {et~al.} 2011, \apjs, 192, 3

\bibitem[{{Paxton} {et~al.}(2013){Paxton}, {Cantiello}, {Arras}, {Bildsten}, {Brown}, {Dotter}, {Mankovich}, {Montgomery}, {Stello}, {Timmes}, \& {Townsend}}]{MESA_II}
{Paxton}, B., {Cantiello}, M., {Arras}, P., {et~al.} 2013, \apjs, 208, 4

\bibitem[{{Paxton} {et~al.}(2015){Paxton}, {Marchant}, {Schwab}, {Bauer}, {Bildsten}, {Cantiello}, {Dessart}, {Farmer}, {Hu}, {Langer}, {Townsend}, {Townsley}, \& {Timmes}}]{MESA_III}
{Paxton}, B., {Marchant}, P., {Schwab}, J., {et~al.} 2015, \apjs, 220, 15

\bibitem[{{Paxton} {et~al.}(2018){Paxton}, {Schwab}, {Bauer}, {Bildsten}, {Blinnikov}, {Duffell}, {Farmer}, {Goldberg}, {Marchant}, {Sorokina}, {Thoul}, {Townsend}, \& {Timmes}}]{MESA_IV}
{Paxton}, B., {Schwab}, J., {Bauer}, E.~B., {et~al.} 2018, \apjs, 234, 34

\bibitem[{{Perley} {et~al.}(2016){Perley}, {Quimby}, {Yan}, {Vreeswijk}, {De Cia}, {Lunnan}, {Gal-Yam}, {Yaron}, {Filippenko}, {Graham}, {Laher}, \& {Nugent}}]{Perley16_hostgalaxies_SLSNe}
{Perley}, D.~A., {Quimby}, R.~M., {Yan}, L., {et~al.} 2016, \apj, 830, 13

\bibitem[{{Perley} {et~al.}(2022){Perley}, {Sollerman}, {Schulze}, {Yao}, {Fremling}, {Gal-Yam}, {Ho}, {Yang}, {Kool}, {Irani}, {Yan}, {Andreoni}, {Baade}, {Bellm}, {Brink}, {Chen}, {Cikota}, {Coughlin}, {Dahiwale}, {Dekany}, {Duev}, {Filippenko}, {Hoeflich}, {Kasliwal}, {Kulkarni}, {Lunnan}, {Masci}, {Maund}, {Medford}, {Riddle}, {Rosnet}, {Shupe}, {Strotjohann}, {Tzanidakis}, \& {Zheng}}]{Perley2022_SN2021csp_Icn}
{Perley}, D.~A., {Sollerman}, J., {Schulze}, S., {et~al.} 2022, \apj, 927, 180

\bibitem[{{Popov} {et~al.}(2025){Popov}, {M{\"u}ller}, \& {Mandel}}]{Popov25_natalkicks_review}
{Popov}, S., {M{\"u}ller}, B., \& {Mandel}, I. 2025, \nar, 101, 101734

\bibitem[{{Pursiainen} {et~al.}(2022){Pursiainen}, {Leloudas}, {Paraskeva}, {Cikota}, {Anderson}, {Angus}, {Brennan}, {Bulla}, {Camacho-I{\~n}iguez}, {Charalampopoulos}, {Chen}, {Delgado Manche{\~n}o}, {Fraser}, {Frohmaier}, {Galbany}, {Guti{\'e}rrez}, {Gromadzki}, {Inserra}, {Maund}, {M{\"u}ller-Bravo}, {Mu{\~n}oz Torres}, {Nicholl}, {Onori}, {Patat}, {Pessi}, {Roy}, {Spyromilio}, {Wiseman}, \& {Young}}]{Pursiainen2022_2018bsz_SLSNI_withH}
{Pursiainen}, M., {Leloudas}, G., {Paraskeva}, E., {et~al.} 2022, \aap, 666, A30

\bibitem[{{Ragosta} {et~al.}(2025){Ragosta}, {Illiano}, {Simongini}, {Rodr{\'\i}guez}, {Imbrogno}, {Piranomonte}, \& {Melandri}}]{Ragosta25_2015ap_CCI}
{Ragosta}, F., {Illiano}, G., {Simongini}, A., {et~al.} 2025, \mnras, 541, 1048

\bibitem[{{Repetto} {et~al.}(2012){Repetto}, {Davies}, \& {Sigurdsson}}]{RepettoDaviesSigurdsson2012_SMBH_kicks_similartoNS}
{Repetto}, S., {Davies}, M.~B., \& {Sigurdsson}, S. 2012, \mnras, 425, 2799

\bibitem[{{Ryder} {et~al.}(2018){Ryder}, {Van Dyk}, {Fox}, {Zapartas}, {de Mink}, {Smith}, {Brunsden}, {Azalee Bostroem}, {Filippenko}, {Shivvers}, \& {Zheng}}]{Ryder2018_companion2001ig}
{Ryder}, S.~D., {Van Dyk}, S.~D., {Fox}, O.~D., {et~al.} 2018, \apj, 856, 83

\bibitem[{{Salpeter}(1955)}]{Salpeter_IMF_55}
{Salpeter}, E.~E. 1955, \apj, 121, 161

\bibitem[{{Sana} {et~al.}(2012){Sana}, {de Mink}, {de Koter}, {Langer}, {Evans}, {Gieles}, {Gosset}, {Izzard}, {Le Bouquin}, \& {Schneider}}]{Sana_massive_stars_binaries}
{Sana}, H., {de Mink}, S.~E., {de Koter}, A., {et~al.} 2012, Science, 337, 444

\bibitem[{{Sana} {et~al.}(2014){Sana}, {Le Bouquin}, {Lacour}, {Berger}, {Duvert}, {Gauchet}, {Norris}, {Olofsson}, {Pickel}, {Zins}, {Absil}, {de Koter}, {Kratter}, {Schnurr}, \& {Zinnecker}}]{Sana14_Ostar_binaries}
{Sana}, H., {Le Bouquin}, J.~B., {Lacour}, S., {et~al.} 2014, \apjs, 215, 15

\bibitem[{{Sana} {et~al.}(2025){Sana}, {Shenar}, {Bodensteiner}, {Britavskiy}, {Langer}, {Lennon}, {Mahy}, {Mandel}, {de Mink}, {Patrick}, {Villase{\~n}or}, {Dirickx}, {Abdul-Masih}, {Almeida}, {Backs}, {Berlanas}, {Bernini-Peron}, {Bowman}, {Bronner}, {Crowther}, {Deshmukh}, {Evans}, {Fabry}, {Gieles}, {Gilkis}, {Gonz{\'a}lez-Tor{\`a}}, {Gr{\"a}fener}, {G{\"o}tberg}, {Hawcroft}, {H{\'e}nault-Brunet}, {Herrero}, {Holgado}, {Izzard}, {de Koter}, {Janssens}, {Johnston}, {Josiek}, {Justham}, {Kalari}, {Klencki}, {Kub{\'a}t}, {Kub{\'a}tov{\'a}}, {Lefever}, {van Loon}, {Ludwig}, {Mackey}, {Ma{\'\i}z Apell{\'a}niz}, {Maravelias}, {Marchant}, {Mazeh}, {Menon}, {Moe}, {Najarro}, {Oskinova}, {Ovadia}, {Pauli}, {Pawlak}, {Ramachandran}, {Renzo}, {Rocha}, {Sander}, {Schneider}, {Schootemeijer}, {Sch{\"o}sser}, {Sch{\"u}rmann}, {Sen}, {Shahaf}, {Sim{\'o}n-D{\'\i}az}, {van Son}, {Stoop}, {Toonen}, {Tramper}, {Valli}, {Vigna-G{\'o}mez}, {Vink}, {Wang}, \& {Willcox}}]{Sana2025_BLOEM_binaryfraction_Z}
{Sana}, H., {Shenar}, T., {Bodensteiner}, J., {et~al.} 2025, Nature Astronomy, 9, 1337

\bibitem[{{Schneider} {et~al.}(2021){Schneider}, {Podsiadlowski}, \& {M{\"u}ller}}]{Schneider_preSNevo_stripped_2021}
{Schneider}, F.~R.~N., {Podsiadlowski}, P., \& {M{\"u}ller}, B. 2021, \aap, 645, A5

\bibitem[{{Schootemeijer} {et~al.}(2018){Schootemeijer}, {G{\"o}tberg}, {de Mink}, {Gies}, \& {Zapartas}}]{Schootemeijer18_PhiPersei}
{Schootemeijer}, A., {G{\"o}tberg}, Y., {de Mink}, S.~E., {Gies}, D., \& {Zapartas}, E. 2018, \aap, 615, A30

\bibitem[{{Schulze} {et~al.}(2018){Schulze}, {Kr{\"u}hler}, {Leloudas}, {Gorosabel}, {Mehner}, {Buchner}, {Kim}, {Ibar}, {Amor{\'\i}n}, {Herrero-Illana}, {Anderson}, {Bauer}, {Christensen}, {de Pasquale}, {de Ugarte Postigo}, {Gallazzi}, {Hjorth}, {Morrell}, {Malesani}, {Sparre}, {Stalder}, {Stark}, {Th{\"o}ne}, \& {Wheeler}}]{Schulze18_hostgalaxies_SLSNe}
{Schulze}, S., {Kr{\"u}hler}, T., {Leloudas}, G., {et~al.} 2018, \mnras, 473, 1258

\bibitem[{{Sch{\"u}rmann} \& {Langer}(2024)}]{SchuermannLanger25_stableRLOF_accretor}
{Sch{\"u}rmann}, C. \& {Langer}, N. 2024, \aap, 691, A174

\bibitem[{{Sch{\"u}rmann} {et~al.}(2025){Sch{\"u}rmann}, {Xu}, {Langer}, {Lennon}, {Kruckow}, {Antoniadis}, {Haberl}, {Herrero}, {Kramer}, {Schootemeijer}, {Shenar}, {Tauris}, \& {Wang}}]{Schuermann25_SMC_pop_combine}
{Sch{\"u}rmann}, C., {Xu}, X.-T., {Langer}, N., {et~al.} 2025, \aap, 704, A219

\bibitem[{{Smith} {et~al.}(2011){Smith}, {Li}, {Filippenko}, \& {Chornock}}]{Smith2011_obsSNfractions}
{Smith}, N., {Li}, W., {Filippenko}, A.~V., \& {Chornock}, R. 2011, \mnras, 412, 1522

\bibitem[{{Sollerman} {et~al.}(2020){Sollerman}, {Fransson}, {Barbarino}, {Fremling}, {Horesh}, {Kool}, {Schulze}, {Sfaradi}, {Yang}, {Bellm}, {Burruss}, {Cunningham}, {De}, {Drake}, {Golkhou}, {Green}, {Kasliwal}, {Kulkarni}, {Kupfer}, {Laher}, {Masci}, {Rodriguez}, {Rusholme}, {Williams}, {Yan}, \& {Zolkower}}]{Sollerman2020_2019tsf_2019oys}
{Sollerman}, J., {Fransson}, C., {Barbarino}, C., {et~al.} 2020, \aap, 643, A79

\bibitem[{{Souropanis} {et~al.}(2026){Souropanis}, {Zapartas}, {Pessi}, {Briel}, {Renzo}, {Guti{\'e}rrez}, {Andrews}, {Gossage}, {Kruckow}, {Liotine}, {Srivastava}, \& {Teng}}]{Souropanis25_SESNe_Metallicity_POSYDON}
{Souropanis}, D., {Zapartas}, E., {Pessi}, T., {et~al.} 2026, \mnras, 545, staf2163

\bibitem[{{Sun} {et~al.}(2022){Sun}, {Maund}, {Crowther}, {Hirai}, {Kashapov}, {Liu}, {Liu}, \& {Zapartas}}]{Sun2022_2019yvr_companion}
{Sun}, N.-C., {Maund}, J.~R., {Crowther}, P.~A., {et~al.} 2022, \mnras, 510, 3701

\bibitem[{{Sun} {et~al.}(2020){Sun}, {Maund}, {Hirai}, {Crowther}, \& {Podsiadlowski}}]{Sun2020_companions2006jc_2015G}
{Sun}, N.-C., {Maund}, J.~R., {Hirai}, R., {Crowther}, P.~A., \& {Podsiadlowski}, P. 2020, \mnras, 491, 6000

\bibitem[{{Taam} \& {Fryxell}(1984)}]{TaamFryxell84_sne_ejecta_companion}
{Taam}, R.~E. \& {Fryxell}, B.~A. 1984, \apj, 279, 166

\bibitem[{{Tauris} \& {Bailes}(1996)}]{TaurisBailes96_ms_pulsars_sesne_lowkicks}
{Tauris}, T.~M. \& {Bailes}, M. 1996, \aap, 315, 432

\bibitem[{{Tauris} {et~al.}(2017){Tauris}, {Kramer}, {Freire}, {Wex}, {Janka}, {Langer}, {Podsiadlowski}, {Bozzo}, {Chaty}, {Kruckow}, \& et~al.}]{Tauris17_DNS_lowkicks}
{Tauris}, T.~M., {Kramer}, M., {Freire}, P.~C.~C., {et~al.} 2017, \apj, 846, 170

\bibitem[{{Tauris} \& {Takens}(1998)}]{TaurisTaken98_binarypostExp_parameters_kicks_impact}
{Tauris}, T.~M. \& {Takens}, R.~J. 1998, \aap, 330, 1047

\bibitem[{{Thorne} \& {$\dot{\mathrm{Z}}$ytkow}(1977)}]{ThorneZytkow_77}
{Thorne}, K.~S. \& {$\dot{\mathrm{Z}}$ytkow}, A.~N. 1977, \apj, 212, 832

\bibitem[{{Valli} {et~al.}(2025){Valli}, {de Mink}, {Justham}, {Callister}, {Johnston}, {Kresse}, {Langer}, {Rubio}, {Vigna-G{\'o}mez}, \& {Wang}}]{Valli25_BeXrayBinaries_kicks}
{Valli}, R., {de Mink}, S.~E., {Justham}, S., {et~al.} 2025, arXiv e-prints, arXiv:2505.08857

\bibitem[{{Vinciguerra} {et~al.}(2020){Vinciguerra}, {Neijssel}, {Vigna-G{\'o}mez}, {Mandel}, {Podsiadlowski}, {Maccarone}, {Nicholl}, {Kingdon}, {Perry}, \& {Salemi}}]{Vinciguerra2020}
{Vinciguerra}, S., {Neijssel}, C.~J., {Vigna-G{\'o}mez}, A., {et~al.} 2020, \mnras, 498, 4705

\bibitem[{{Wang} {et~al.}(2016){Wang}, {Jia}, \& {Li}}]{Wang_CEE}
{Wang}, C., {Jia}, K., \& {Li}, X.-D. 2016, Research in Astronomy and Astrophysics, 16, 126

\bibitem[{{Wang} {et~al.}(2026){Wang}, {Lau}, {Li}, {Langer}, {de Mink}, {Valli}, {Justham}, {Xu}, {Klencki}, \& {Ryu}}]{Wang26_accretion_and_criticalrotation}
{Wang}, C., {Lau}, M. Y.~M., {Li}, X.-D., {et~al.} 2026, \aap, 707, A14

\bibitem[{{Webbink}(1984)}]{Webbing_alphalambda}
{Webbink}, R.~F. 1984, \apj, 277, 355

\bibitem[{{West} {et~al.}(2023){West}, {Lunnan}, {Omand}, {Kangas}, {Schulze}, {Strotjohann}, {Yang}, {Fransson}, {Sollerman}, {Perley}, {Yan}, {Chen}, {Chen}, {Taggart}, {Fremling}, {Bloom}, {Drake}, {Graham}, {Kasliwal}, {Laher}, {Medford}, {Neill}, {Riddle}, \& {Shupe}}]{West_2020qlb_SLSNI_oscillations}
{West}, S.~L., {Lunnan}, R., {Omand}, C.~M.~B., {et~al.} 2023, \aap, 670, A7

\bibitem[{{Wheeler} {et~al.}(1975){Wheeler}, {Lecar}, \& {McKee}}]{Wheeler75_SNe_companion_ablation}
{Wheeler}, J.~C., {Lecar}, M., \& {McKee}, C.~F. 1975, \apj, 200, 145

\bibitem[{{Williamson} {et~al.}(2021){Williamson}, {Kerzendorf}, \& {Modjaz}}]{Williamson21_1994I_Ic_with_He}
{Williamson}, M., {Kerzendorf}, W., \& {Modjaz}, M. 2021, \apj, 908, 150

\bibitem[{{Xu} {et~al.}(2025){Xu}, {Sch{\"u}rmann}, {Langer}, {Wang}, {Schootemeijer}, {Shenar}, {Ercolino}, {Haberl}, {Hastings}, {Jin}, {Kramer}, {Lennon}, {Marchant}, {Sen}, {Tauris}, \& {de Mink}}]{Xu25_SMC_pop_mesa}
{Xu}, X.-T., {Sch{\"u}rmann}, C., {Langer}, N., {et~al.} 2025, \aap, 704, A218

\bibitem[{{Yan} {et~al.}(2017){Yan}, {Lunnan}, {Perley}, {Gal-Yam}, {Yaron}, {Roy}, {Quimby}, {Sollerman}, {Fremling}, {Leloudas}, {Cenko}, {Vreeswijk}, {Graham}, {Howell}, {De Cia}, {Ofek}, {Nugent}, {Kulkarni}, {Hosseinzadeh}, {Masci}, {McCully}, {Rebbapragada}, \& {Wo{\'z}niak}}]{Yan2017_SLSNeI_withHalpha_and_oscillations}
{Yan}, L., {Lunnan}, R., {Perley}, D.~A., {et~al.} 2017, \apj, 848, 6

\bibitem[{{Yoon}(2015)}]{Yoon15_Ibc_Hecontent}
{Yoon}, S.-C. 2015, \pasa, 32, e015

\bibitem[{{Zapartas} {et~al.}(2019){Zapartas}, {de Mink}, {Justham}, {Smith}, {de Koter}, {Renzo}, {Arcavi}, {Farmer}, {G{\"o}tberg}, \& {Toonen}}]{Zapartas19_howmany_Hrich_Sne_from_binaries}
{Zapartas}, E., {de Mink}, S.~E., {Justham}, S., {et~al.} 2019, \aap, 631, A5

\bibitem[{{Zapartas} {et~al.}(2026){Zapartas}, {Fox}, {Su}, {Souropanis}, {Drout}, {Rocha}, {van Dyk}, {Williams}, {Briel}, {Renzo}, {Andrews}, {Fragos}, {Gossage}, {Kruckow}, {Liotine}, {Ryder}, {Srivastava}, \& {Teng}}]{Zapartas25_demographics_companions_of_SESNe}
{Zapartas}, E., {Fox}, O.~D., {Su}, J., {et~al.} 2026, \mnras, 546, staf2208

\bibitem[{{Zenati} {et~al.}(2025){Zenati}, {Wang}, {Bobrick}, {DeMarchi}, {Glanz}, {Rozner}, {Jencson}, {Rest}, {Metzger}, {Margutti}, {Gomez}, {Smith}, {Toonen}, {Bright}, {Norman}, {Foley}, {Gagliano}, {Krolik}, {Smartt}, {Villar}, {Narayan}, {Fox}, {Auchettl}, {Brethauer}, {Clocchiatti}, {Coelln}, {Coppejans}, {Dimitriadis}, {Dorozsmai}, {Drout}, {Jacobson-Galan}, {Gao}, {Ridden-Harper}, {Kilpatrick}, {Laskar}, {Matthews}, {Rest}, {Smith}, {Stauffer}, {Stroh}, {Strolger}, {Terreran}, {Pierel}, \& {Piro}}]{Zenati2025_SN2019tsf_three_peaked_Ib}
{Zenati}, Y., {Wang}, Q., {Bobrick}, A., {et~al.} 2025, \apj, 992, 9

\bibitem[{{Zhang} {et~al.}(2025){Zhang}, {Wang}, \& {Ji}}]{Zhang25_2022jli_gammarays}
{Zhang}, P., {Wang}, Z., \& {Ji}, S. 2025, arXiv e-prints, arXiv:2512.09223

\bibitem[{{Zhu} {et~al.}(2026){Zhu}, {Nordin}, {de O{\~n}a-Wilhelmi}, \& {Ercolino}}]{Zhu2026_binariesCCSNeZTF}
{Zhu}, S.~J., {Nordin}, J., {de O{\~n}a-Wilhelmi}, E., \& {Ercolino}, A. 2026, arXiv e-prints, arXiv:2608.01994

\bibitem[{{Zimmerman} {et~al.}(2024){Zimmerman}, {Irani}, {Chen}, {Gal-Yam}, {Schulze}, {Perley}, {Sollerman}, {Filippenko}, {Shenar}, {Yaron}, {Shahaf}, {Bruch}, {Ofek}, {De Cia}, {Brink}, {Yang}, {Vasylyev}, {Ben Ami}, {Aubert}, {Badash}, {Bloom}, {Brown}, {De}, {Dimitriadis}, {Fransson}, {Fremling}, {Hinds}, {Horesh}, {Johansson}, {Kasliwal}, {Kulkarni}, {Kushnir}, {Martin}, {Matuzewski}, {McGurk}, {Miller}, {Morag}, {Neil}, {Nugent}, {Post}, {Prusinski}, {Qin}, {Raichoor}, {Riddle}, {Rowe}, {Rusholme}, {Sfaradi}, {Sjoberg}, {Soumagnac}, {Stein}, {Strotjohann}, {Terwel}, {Wasserman}, {Wise}, {Wold}, {Yan}, \& {Zhang}}]{Ziimerman2024_2023ixf}
{Zimmerman}, E.~A., {Irani}, I., {Chen}, P., {et~al.} 2024, \nat, 627, 759

\end{thebibliography}

\begin{appendix} 
\FloatBarrier

\section{Calibration of the reference model}\label{sec:appendix:OB_cal}

We compare our population models to \citet{Schuermann25_SMC_pop_combine}, who used the rapid population-synthesis code ComBinE \citep{Kruckow18_COMBINE} to reproduce the number of Be/X-ray and Wolf-Rayet binaries observed in the SMC.
They report that, following core collapse of the primary star with an OB star companion, $36\%$ of the companions orbit a NS, $35\%$ are disrupted by the supernova, and $29\%$ form a BH companion. We denote these estimates as $f_\mathrm{NS,bound}=36\%$, $f_\mathrm{NS,unbound}=35\%$ and $f_\mathrm{BH}=29\%$. Comparing these numbers to our models allows us to constrain the natal kick distribution, which strongly affects the number and features of supernovae exhibiting CCI (see Sect.\,\ref{sec:results:multiple:kick}).

\begin{figure}
    \centering
    \includegraphics[width=\linewidth]{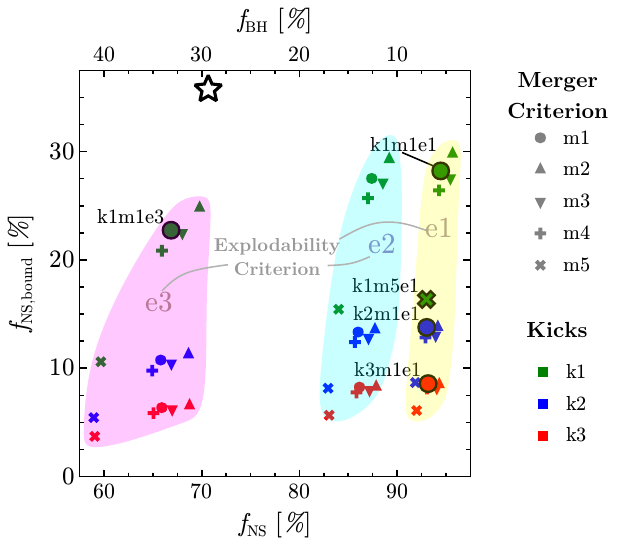}
    \caption{Fraction of binaries in our population models where the primary star explodes and produces a NS $f_\mathrm{NS}$ ($x$-axis), and those which produce a NS that remains bound to the companion $f_\mathrm{NS,bound}$ ($y$-axis). Both fractions are normalized to all systems where the primary undergoes core collapse: $f_\mathrm{NS}+f_\mathrm{BH}=1$ with $f_\mathrm{NS}=f_\mathrm{NS,bound}+f_\mathrm{NS,unbound}$. Each scatter point is a population model, where colors indicate the kick prescription (\texttt{k1}–\texttt{k3}), markers the merger criterion (m1–m5), and shaded regions the explodability criterion (e1–e3; see Table\,\ref{tab:acronyms}) adopted. The five models discussed in the main text are highlighted and labeled. The star marker is the results from the fiducial model of \cite{Schuermann25_SMC_pop_combine}.}
    \label{fig:OB_companions}
\end{figure}

Using $f_\mathrm{NS,bound}$ as our calibration diagnostic, we find that the population models employing the \Valli\ kicks yield the closest agreement with \cite{Schuermann25_SMC_pop_combine} ($f_\mathrm{NS,bound}=11-30\%$; Fig.\,\ref{fig:OB_companions}). The \COMBINE\ and \DM\ kick prescriptions produce smaller $f_\mathrm{NS,bound}$ ($5-14\%$ and $4-9\%$ respectively). Although \citet{Schuermann25_SMC_pop_combine} adopt the \COMBINE\ kicks, our models employing them predict smaller bound fractions than \cite{Schuermann25_SMC_pop_combine}. This reflects additional differences in our models, as outlined below.

First, \cite{Schuermann25_SMC_pop_combine} predicts population numbers by accounting for stellar lifetimes, which preferentially increases the contribution of low-mass accretors (i.e. binaries with lower $\qi$) than high-mass ones (high $\qi$) that share the same birth probability. Furthermore, binaries with low $\qi$ are more likely to remain bound after the primary’s explosion because they are typically tighter at core collapse. We therefore expect systematically fewer bound NS binaries than \citet{Schuermann25_SMC_pop_combine}.

A similar conclusion can be draw when comparing the different metallicity and wind mass-loss recipes adopted in \cite{Schuermann25_SMC_pop_combine}. They use SMC-metallicity models from \cite{Brott2011}, whereas we use solar-metallicity models from \cite{Jin2024_boron,Jin25_BonnGal}, which also feature higher wind mass-loss rates, especially for stripped and partially stripped stars. These two factors lead our binary models to overall stronger orbital widening than in ComBinE and thus to a smaller $f_\mathrm{NS,bound}$. On the other hand, this trend is partly counteracted by the higher mass-transfer efficiency adopted by \cite{Schuermann25_SMC_pop_combine}  which reduces angular-momentum losses and can also widen orbits relative to our low accretion-efficiency models. 

Finally, \cite{Schuermann25_SMC_pop_combine} adopt a different explodability criterion than in any of our population models, which can shift systems between producing NSs or BHs, thus affecting $f_\mathrm{NS,bound}$. However, this effect in our population models is secondary, compared to that of different kick distributions (Fig.\,\ref{fig:OB_companions}). 

Because some of the aforementioned effects act in opposite directions and we cannot robustly quantify their net impact, this comparison does not permit a strong calibration of natal kicks. We therefore adopt the \Valli\ kicks as a pragmatic reference choice since they yield the closest agreement with the estimate of $f_\mathrm{NS,bound}$ in \citet{Schuermann25_SMC_pop_combine}.

\section{Parameter space for supernovae exhibiting periodic CCI}\label{app:parameter_space}

CCI requires substantial radial expansion of the companion following ECI (i.e. large $R_\mathrm{max}$) that must persist long enough so the compact object can penetrate its inflated envelope multiple times (i.e. large $\tauinfl$). Both quantities scale with the energy injected to the companion by the supernova explosion $E_\mathrm{heat}$ (Eqs.\,\ref{eq:Rmax}, \ref{eq:tau_ECI}), which is also higher in the binaries which are tighter at the moment of core collapse. As a result, the most strongly inflated companions, which also remain inflated for the longest time, are also less likely to break up following the supernova. 

\begin{figure}
    \centering
    \includegraphics[width=\linewidth]{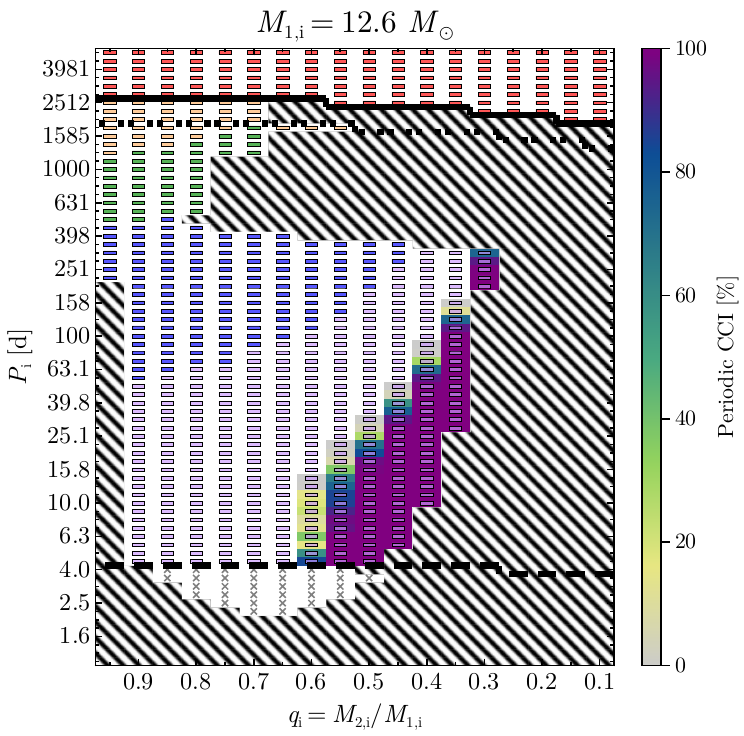}
    \includegraphics[width=\linewidth]{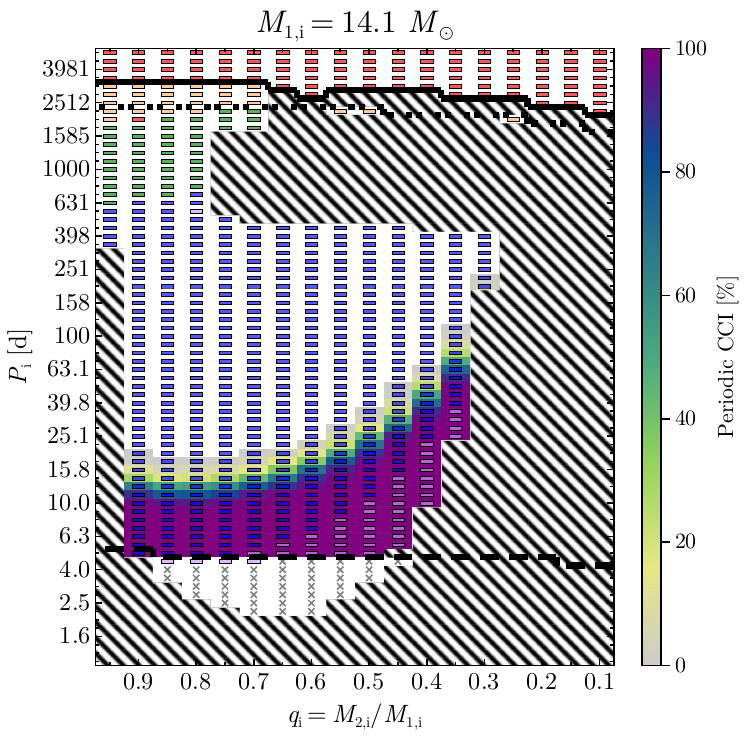}
    \caption{The $\log P_\mathrm{i}-\qi$ diagrams showing the probability that the first supernova in each binary will exhibit periodic CCI (see colorbar) within the reference population model for binaries with $M_\mathrm{1,i}=12.6\Msun$ (top) and $14.5\Msun$ (bottom). Hatched regions indicate evolutionary models that merge before the first supernova occurs. Rectangular scatters are colored corresponding to the supernova type of the primary star (blue for \Type Ibc, violet for \Type Ibn, green for \Type IIb, red for \Type IIP/L and orange for \Type IIn), while gray crosses indicate those that become a white dwarf (gray).}
    \label{fig:logPq}
\end{figure}

Figure\,\ref{fig:logPq} shows the birth parameter space of binaries that produce supernovae with periodic CCI in the reference population model. As expected, shorter initial orbital periods $P_\mathrm{i}$ are more likely to produce supernovae with periodic CCI. We also find that these supernovae can be produced in binaries with low initial mass ratios $\qi$ and higher $P_\mathrm{i}$, because RLOF and the associated angular-momentum losses can lead these systems to tighter post-RLOF orbital separations.

Binary systems with low $\qi$ and $P_\mathrm{i}$ are more likely to undergo \Case BB RLOF \citep{Ercolino_HpoorInteractingSNe}, which reduces the ejecta mass. Under the kick prescriptions of \COMBINE \ or \Valli, the newly born NSs in these systems receive smaller kicks, making them more likely to remain bound to the binary companion following the primary's supernova and thereby increasing the fraction of supernovae with periodic CCI. However, \Case BB RLOF widens the orbit, especially at high $\qi$, so high $\qi$ systems are less likely to produce supernovae with periodic CCI (Fig.\,\ref{fig:logPq}). 

A larger companion radius can also help achieve periodic CCI by intercepting more energy during the explosion (Eq.\,\ref{eq:OmegaEff}-\ref{eq:Eheat}). This is achieved in binary models where the the companion is more evolved in the main-sequence by the time the progenitor star undergoes core collapse, which requires a high $\qi$. This trend appears as a slight increase in the likelihood of periodic CCI in the models with $\qi=0.90$ relative to those with $\qi=0.80$ (see bottom panels in Figs.\,\ref{fig:logPq}, \ref{fig:logPq_diffpop_1},\ref{fig:logPq_diffpop_2}). While this effect has little population-wide impact, it could become more significant with binary models assuming higher accretion efficiencies (Sect.\,\ref{sec:disc:models}).  

\FloatBarrier

\section{Observing the inflated companion following the primary's supernova}\label{sec:appendix:observability}
To determine the observability of the companions to SN2022jli, we convert the companion models' luminosity $L_\mathrm{max}$ and effective temperature $T_\mathrm{eff}$ into $J$-band magnitudes using the bolometric corrections available from MIST \citep{Dotter16_MIST, Choi16_MIST}. In particular, we constructed linear-interpolation maps of their solar-metallicity tables as a function of effective temperature $T_\mathrm{eff}$, surface gravity $\log g$ and visual extinction $A_V$ for a given instrument and filter.

The focus to SN2022jli is motivated by the fact that the companion may still be inflated at the time of writing (Sect.\,\ref{sec:observations:2022jli}), and its distance is smaller ($22.5\,\mathrm{Mpc}$, \citealt{Mould2000_HST_extragalactic_distances}) than the other transients with periodic CCI discussed in this work. For its host galaxy, NGC\,157, we adopt a reddening of $E(B-V)=0.25$ \citep{Chen2024_SN2022jli_binary} and we assume $R_V=3.1$ \citep{Cardelli1989_extinction}, yielding $A_V=0.775$. The observations conducted by \cite{Grosbol12_straformation_GDgalaxies} were able to identify stars close to the location of SN\,2022jli with magnitudes in the VLT's HAWK-I $J$-band of around $\sim21-23$ with about 4-minute exposure time. 

\begin{figure}
    \centering
    \includegraphics[width=\linewidth]{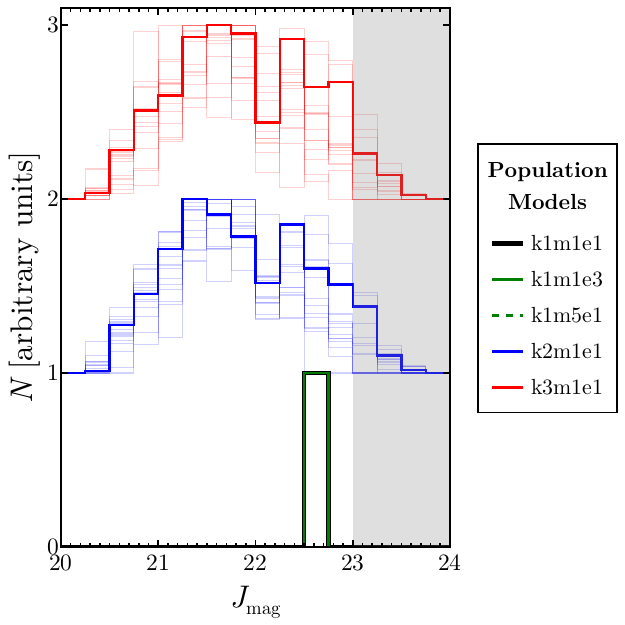}
    \caption{Apparent magnitude histograms of the companion star of SN2022jli during its inflated phase, as observed through the VLT's VISTA/HAWK-I $J$-band filter. Multiple histograms refer to different population models (see Fig.\,\ref{fig:distr}), shifted to different heights for different kick prescriptions. The range below the detection limit observed through the VLT's HAWK-I instrument \citep{Grosbol12_straformation_GDgalaxies} is grayed out.}
    \label{fig:Jband}
\end{figure}

\begin{figure}
    \centering
\begin{tikzpicture}
\node[anchor=south west, inner sep=0] (img){\includegraphics[width=\linewidth, trim=0 0 0 0, clip]{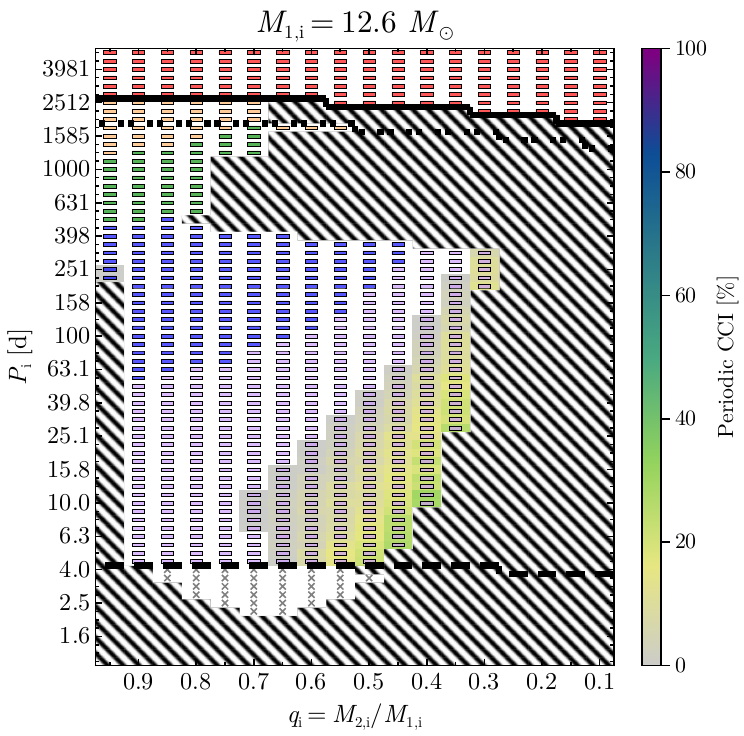}};
\begin{scope}[x={(img.south east)}, y={(img.north west)}]
\node[anchor=north west] at (0.37,1.06) {\textbf{Model \texttt{k3m1e1}}};
\end{scope}
\end{tikzpicture}
{\includegraphics[width=\linewidth, trim=0 0 0 0, clip]{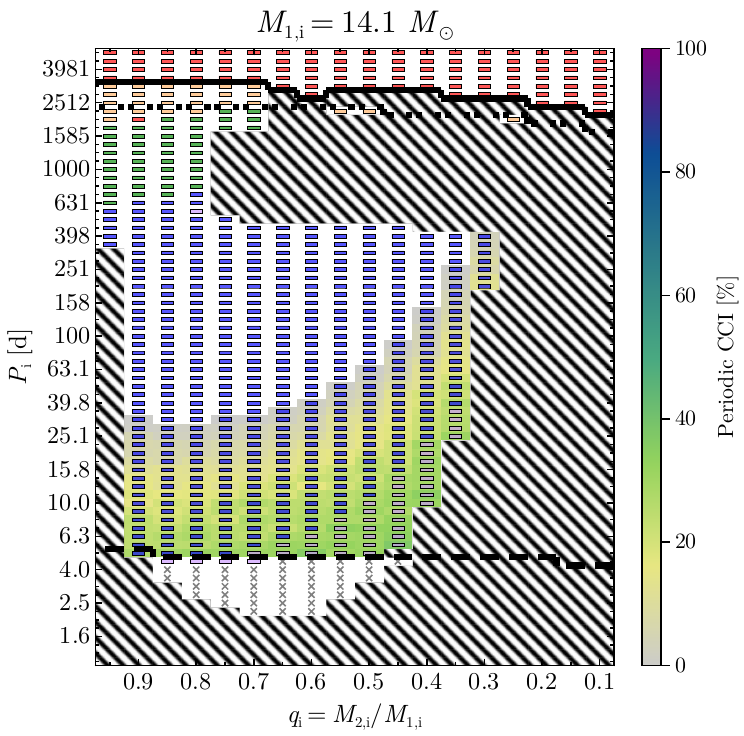}}
\caption{Same as Fig.\,\ref{fig:logPq}, for model \texttt{k3m1e1}.} \label{fig:logPq_diffpop_1}
\end{figure}

\begin{figure}
    \centering
\begin{tikzpicture}
\node[anchor=south west, inner sep=0] (img){\includegraphics[width=\linewidth, trim=0 0 0 0, clip]{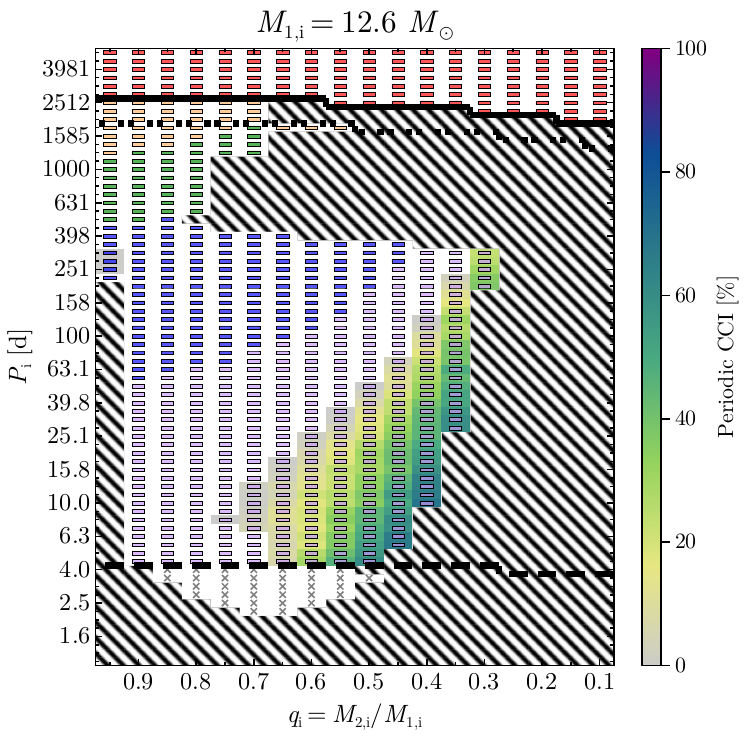}};
\begin{scope}[x={(img.south east)}, y={(img.north west)}]
\node[anchor=north west] at (0.37,1.06) {\textbf{Model \texttt{k2m1e1}}};
\end{scope}
\end{tikzpicture}
{\includegraphics[width=\linewidth, trim=0 0 0 0, clip]{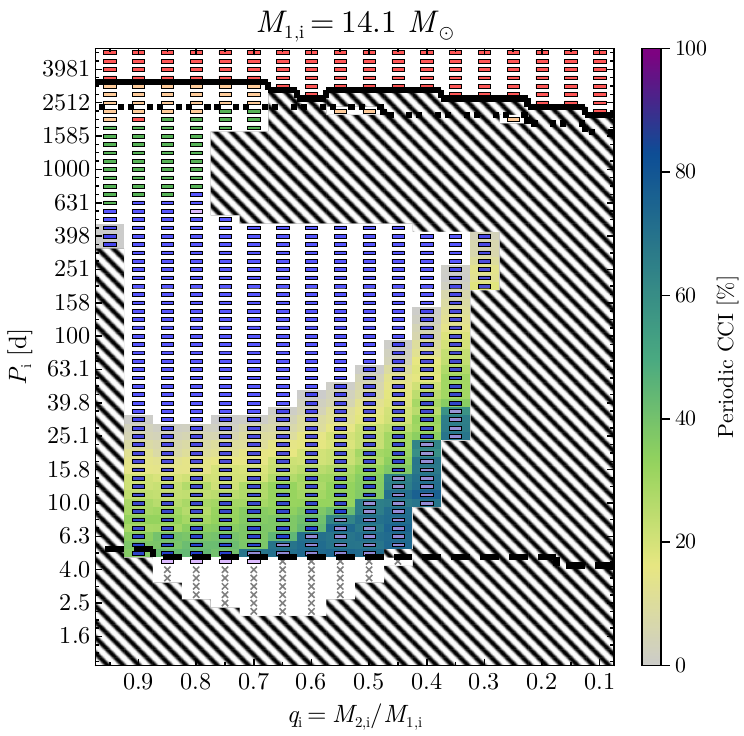}}
\caption{Same as Fig.\,\ref{fig:logPq}, for model \texttt{k2m1e1}. \label{fig:logPq_diffpop_2}}
\end{figure}

We compare this magnitude range to that our models would have when observed via the VISTA's $J$ filter, which is the same as that used in HAWK-I. In our reference population model, SN\,2022jli's companion's $J$-band magnitude would be $\sim22.6$, while other population models (that find one or more binary models compatible with the constrains of SN2022jli) yield a range of magnitudes $20.5-23.5$ (Fig.\,\ref{fig:Jband}). We follow the same procedure to make predictions on the observable magnitude with JWST and HST (Table\,\ref{tab:2022jli_HST_JWST}).
All population models that find compatible binary evolution models explaining the features of SN2022jli predict that the companion is likely detectable with instruments such as the VLT's HAWK-I. The magnitude limit of $\sim23$  can also be significantly improved as it was set with a 4-minute observation.  Longer exposures will allow to set a much deeper limit.  

\FloatBarrier

\section{Additional Data}

\begin{figure}
    \centering
\begin{tikzpicture}
\node[anchor=south west, inner sep=0] (img){\includegraphics[width=\linewidth, trim=0 0 0 0, clip]{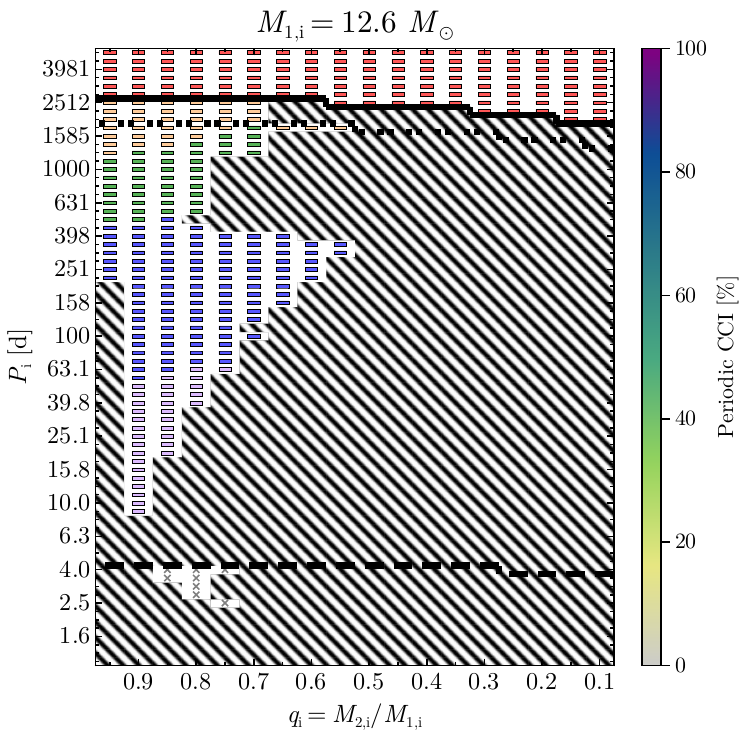}};
\begin{scope}[x={(img.south east)}, y={(img.north west)}]
\node[anchor=north west] at (0.37,1.06) {\textbf{Model \texttt{k1m5e1}}};
\end{scope}
\end{tikzpicture}
{\includegraphics[width=\linewidth, trim=0 0 0 0, clip]{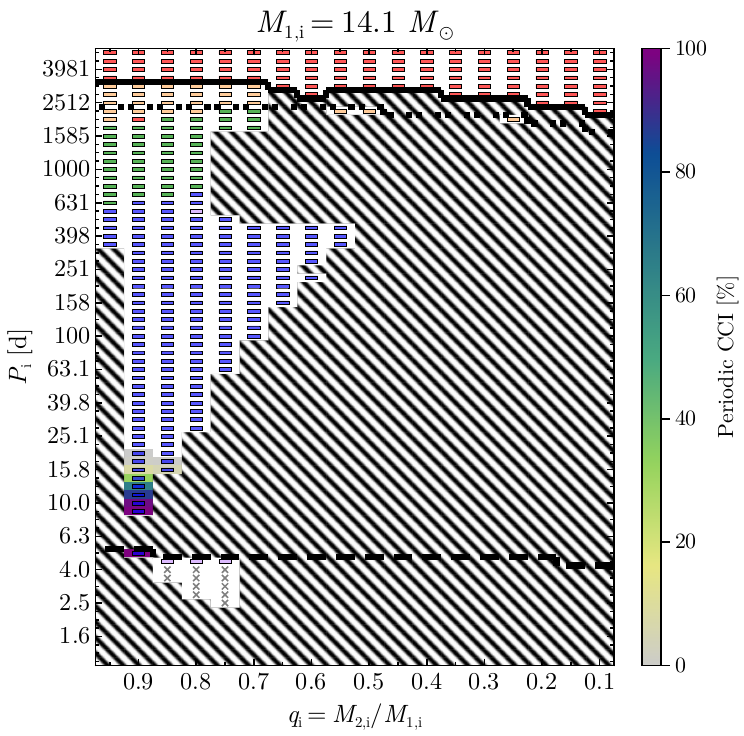}}
\caption{Same as Fig.\,\ref{fig:logPq}, for model \texttt{k1m5e1}.} \label{fig:logPq_diffpop_3}
\end{figure}

\begin{table}
\centering
\caption{Extended version of Table\,\ref{tab:inferred_data_2022jli}, showing the inferred parameters from other population models to \SN{2022jli} (top), \SN{2015ap} (middle) and \SN{2022esa} (bottom).}\label{tab:2022jli_ext}
\resizebox{\linewidth}{!}{
\begin{tabular}{r||cccc}
\textbf{SN2022jli}&\texttt{k1m1e3} & \texttt{k1m5e1} & \texttt{k2m1e1} & \texttt{k3m1e1} \\\hline\hline
$M_\mathrm{1,i} \ [\Msun]$ & $28.2$ & $/$ & $28.2-39.8$ & $28.2-39.8$  \\
$M_\mathrm{2,i}  \ [\Msun]$ & $7.1$ &$/$ &$5.6-29.9$ &$5.6-29.0$ \\
$\qi=M_\mathrm{2,i}/M_\mathrm{1,i}$ & $0.25$ & $/$ & $0.20-0.80$ & $0.20-0.80$  \\
$P_\mathrm{i} \ [\days]$ & $28.2$ & $/$ & $4.0-239.0$ & $3.7-177.8$ \\
Binary models &1 &0 &278 &228\\\hline
$\Mej  \ [\Msun]$ & $3.9$ & $/$ & $3.9-5.3$ & $3.9-5.2$   \\
$M_\mathrm{ej,He}  \ [\Msun]$ & $0.17$ & $/$ & $0.17-0.27$ & $0.17-0.25$ \\
$Y_\mathrm{pre-SN}$ & $0.30$ & $/$ & $0.20-0.46$ & $0.20-0.44$  \\\hline
$M_2  \ [\Msun]$ & $7.4$ & $/$ & $5.8-32.6$ & $5.9-32.6$ \\
$\tauinfl \ [\yr]$ & $9.8$ & $/$ & $1.6-8.0$ & $1.6-7.7$  \\
$L_\mathrm{max} \ [10^{38}\ergs]$ & $3.5$ & $/$ & $2.7-37.8$ & $2.7-37.9$  \\
$T_\mathrm{eff} \ [\mathrm{kK}]$ & $4.0$ & $/$ & $4.1-6.1$ & $4.1-6.2$   \\
$e_\mathrm{post-SN}$ & $0.52$ & $/$ & $0.51-0.83$ & $0.51-0.83$   \\
 \multicolumn{1}{l}{ } & 
\multicolumn{1}{c}{ } & 
\multicolumn{1}{c}{ }& 
\multicolumn{1}{c}{ }\\
\textbf{SN2015ap} &\texttt{k1m1e3} & \texttt{k1m5e1} & \texttt{k2m1e1} & \texttt{k3m1e1} \\\hline\hline
$M_\mathrm{1,i} \ [\Msun]$ & $15.9-31.6$ & $22.4-44.7$ & $14.1-25.1$ &$14.1-25.1$ \\
$M_\mathrm{2,i}  \ [\Msun]$ &$7.1-15.1$ &$13.4-14.6$ &$7.1-33.8$ & $5.6-16.0$  \\
$\qi=M_\mathrm{2,i}/M_\mathrm{1,i}$ & $0.35-0.60$ & $0.30-0.65$ & $0.35-0.85$ & $0.35-0.85$  \\
$P_\mathrm{i} \ [\days]$ & $2.8-10.0$ & $2.5-7.9$ & $2.5-39.8$ & $2.5-35.5$  \\
Binary models &76 &4 &389 &372   \\\hline
$\Mej  \ [\Msun]$ &  $1.8-2.7$ & $2.2-2.5$ & $1.6-2.7$ & $1.6-2.7$  \\
$M_\mathrm{ej,He}  \ [\Msun]$ &  $1.20$ & $1.20$ & $1.1-1.2$ & $1.1-1.2$  \\
$Y_\mathrm{pre-SN}$ &   $0.98$ & $0.98$ & $0.98$ & $0.98$   \\\hline
$M_2  \ [\Msun]$ &  $7.4-17.8$ & $14.7-18.5$ & $5.7-22.4$ & $5.7-21.7$  \\
$\tauinfl \ [\yr]$ &   $4.6-15.6$ & $5.9-7.2$ & $2.5-11.6$ & $2.5-10.2$  \\
$L_\mathrm{max} \ [10^{38}\ergs]$ &  $3.5-11.2$ & $8.4-11.9$ & $2.6-16.3$ & $2.6-15.5$  \\
$T_\mathrm{eff} \ [\mathrm{kK}]$  & $3.7-4.9$ & $4.7-4.8$ & $3.9-5.2$ & $4.0-5.2$   \\
$e_\mathrm{post-SN}$ &  $0.18-0.41$ & $0.16-0.25$ & $0.12-0.79$ & $0.15-0.80$  \\
 \multicolumn{1}{l}{ } & 
\multicolumn{1}{c}{ } & 
\multicolumn{1}{c}{ }& 
\multicolumn{1}{c}{ }\\
\textbf{SN2022esa}&\texttt{k1m1e3} & \texttt{k1m5e1} & \texttt{k2m1e1} & \texttt{k3m1e1} \\\hline\hline
$M_\mathrm{1,i} \ [\Msun]$ & $28.2-44.7$ & $28.2-44.7$ & $28.2-50.1$ & $21.2-50.1$ \\
$M_\mathrm{2,i}  \ [\Msun]$ & $5.6-29.0$ & $11.1-33.5$ & $7.1-33.8$ & $6.3-33.8$ \\
$\qi=M_\mathrm{2,i}/M_\mathrm{1,i}$ &   $0.20-0.80$ & $0.35-0.85$ & $0.20-0.85$ & $0.20-0.85$ \\
$P_\mathrm{i} \ [\days]$ &  $5.6-89.1$ & $5.6-12.6$ & $4.47-141.3$ & $4.5-177.8$ \\
Binary models &61 &212 &1025 &1023 \\\hline
$\Mej  \ [\Msun]$ & $3.9-4.4$ & $3.8-5.1$ & $3.8-5.9$ & $3.8-5.9$ \\
$M_\mathrm{ej,He}  \ [\Msun]$ & $0.17-0.30$ & $0.17-0.28$ & $0.17-0.24$ & $0.17-0.24$ \\
$Y_\mathrm{pre-SN}$ & $0.22-0.50$ & $0.21-0.49$ & $0.20-0.43$ & $0.20-0.43$ \\\hline
$M_2  \ [\Msun]$ & $5.9-29.2$ & $11.4-34.5$ & $7.3-34.2$ & $6.6-34.1$ \\
$\tauinfl \ [\yr]$ &  $1.1-8.4$ & $1.3-7.2$ & $0.6-6.7$ & $0.6-6.8$ \\
$L_\mathrm{max} \ [10^{38}\ergs]$  & $2.7-28.2$ & $6.0-45.0$ & $3.4-43.5$ & $3.1-43.3$ \\
$T_\mathrm{eff} \ [\mathrm{kK}]$ &  $4.0-6.6$ & $5.1-6.5$ & $4.8-7.2$ & $4.8-7.2$ \\
$e_\mathrm{post-SN}$ & $0.26-0.70$ & $0.25-0.49$ & $0.15-0.85$ & $0.16-0.85$ \\
\end{tabular}
}
\end{table}

\begin{table}[!htp]\centering\small
\caption{Predicted apparent magnitudes of the companion star of SN2022jli as observed through various photometric filters with the James Webb Space Telescope (JWST) NIRCam and the Hubble Space Telescope (HST) Wide-Field Camera 3.}\label{tab:2022jli_HST_JWST}
\resizebox{\linewidth}{!}{
\begin{tabular}{r||ccccc}
Filter &  \multicolumn{5}{c}{Population models} \\[0.5em]
JWST &\texttt{k1m1e1} &\texttt{k1m1e3} &\texttt{k1m5e1} & \texttt{k2m1e1} &\texttt{k3m1e1} \\\hline\hline
F070W &$24.5-24.6$ & $24.5$ &/ &$21.7$ &$21.7-24.6$ \\
F090W & $23.7$ & $23.7$ &/ & $21.2-24.0$ & $21.3-23.9$ \\
F115W & $22.9      $ & $22.9$ &/ & $20.9-23.4$ & $21.0-23.3$ \\
F150W & $22.1$ & $22.1$ &/ & $20.6-22.8$ & $20.7-22.7$ \\
F200W & $21.6-21.7$ & $21.7$ &/ & $20.5-22.6$ & $20.5-22.4$ \\
F277W & $21.6-21.7$ & $21.7$ &/ & $20.4-22.5$ & $20.4-22.4$ \\
F356W & $21.5$ & $21.5$ &/ & $20.3-22.4$ & $20.4-22.3$ \\
F444W & $21.6$ & $21.6$ &/ & $20.3-22.5$ & $20.3-22.3$ \\
\multicolumn{6}{c}{ } \\
HST &\texttt{k1m1e1} &\texttt{k1m1e3} &\texttt{k1m5e1} & \texttt{k2m1e1} &\texttt{k3m1e1} \\\hline\hline
F275W & $34.4$ & $34.4$ & / & $26.0-34.6$ & $26.1-34.2$ \\
F336W & $31.0-31.2$ & $31.0$ & / & $24.2-31.0$ & $24.2-30.3$ \\
F475W & $26.9-27.0$ & $26.9$ & / & $22.8-27.1$ & $22.7-26.7$ \\
F555W & $26.0-26.1$ & $26.0$ & / & $22.4-26.3$ & $22.4-25.9$ \\
F606W & $25.4$ & $25.4$ & / & $22.1-25.6$ & $22.1-25.3$ \\
F814W & $24.0$ & $24.0$ & / & $21.4-24.3$ & $21.5-24.2$ \\
\end{tabular}}
\end{table}

\FloatBarrier

\end{appendix}

\end{document}